\documentclass[aps,pre,preprint,superscriptaddress, showkeys]{revtex4-1}
\usepackage{amssymb}
\usepackage{amsfonts}
\usepackage{amsmath}
\usepackage{graphicx}
\usepackage[caption=false]{subfig}
\usepackage{morefloats}
\usepackage{color}
\usepackage{epstopdf}

\begin{document}

\title{On Synchronization of Interdependent Networks}
\author{J. Martin-Hernandez}

\affiliation{Faculty of Electrical Engineering, Mathematics and
Computer Science, P.O Box 5031, 2600 GA Delft, The Netherlands}
\affiliation{ENEA Centro Ricerche Casaccia via Anguillarese 301, I-00123 Roma
(RM) , Italy}

\author{H. Wang }
\affiliation{Faculty of Electrical Engineering, Mathematics and
Computer Science, P.O Box 5031, 2600 GA Delft, The Netherlands}

\author{P. Van Mieghem }
\affiliation{Faculty of Electrical Engineering, Mathematics and
Computer Science, P.O Box 5031, 2600 GA Delft, The Netherlands}

\author{G. D'Agostino}
\affiliation{ENEA Centro Ricerche Casaccia via Anguillarese 301, I-00123 Roma
(RM) , Italy}

\date{\today}

\begin{abstract}

It is well-known that the synchronization of diffusively-coupled systems on
networks strongly depends on the network topology. In particular, the so-called
algebraic connectivity $\mu_{N-1}$, or the smallest non-zero eigenvalue of the
discrete Laplacian operator plays a crucial role on synchronization, graph
partitioning, and network robustness. In our study, synchronization is placed in
the general context of networks-of-networks, where single network models are
replaced by a more realistic hierarchy of interdependent networks. The present
work shows, analytically and numerically, how the algebraic connectivity
experiences sharp transitions after the addition of sufficient links among
interdependent networks.

\end{abstract}


\maketitle

\section{Introduction}\label{sec_intro}

In the past decades, there has been a significant advance in understanding the
structure and function of complex networks \cite{A.L.Barabasi1999, WatStr98}.
Mathematical models of networks are now widely used to describe a broad range of
complex systems, from networks of human contacts to interactions amongst
proteins. In particular, synchronization as an emerging phenomenon of a
population of dynamically interacting units has always fascinated humans. The
scientific interest in synchronization of coupled oscillators can be traced back
to the work by Christiaan Huygens on \textit{``An odd kind sympathy''}, between
coupled pendulum clocks \cite{Huygens1673}, where he noticed that two pendulum
clocks mounted on the same frame will synchronize after some time.
Synchronization phenomena and processes are ubiquitous in nature and play a
vital role within various contexts in biology, chemistry, ecology, sociology,
and technology \cite{Bergen1981}, extending topics as epidemic spread
\cite{VanMieghem2011} and coupled oscillators \cite{Strogatz2000, Jadbabaie04,
abprs05}. To date, the problem of how the structural properties of a network
influence the performance and stability of the fully synchronized states of the
network have been extensively investigated and discussed, both numerically and
theoretically \cite{Atay06,Chen12,Dorfler12,Wang2002}.
 
There exist many different definitions for synchronization (most of which are
related). However, in the present paper, we will employ a definition based on
the following equations:

\begin{equation}
\frac{d s_i}{d t} \approx \sum_{j \in N_i}(s_i(t)-s_j(t)) \approx
\sum\nolimits_{j = 1}^N {{Q_{ij}}{s_j}(t)};
\label{eq_synch0}
\end{equation}

\noindent where $s_i$ represents the (relative) deviation of the $i-th$
component state from its equilibrium, $N_i$ its neighbors, and $Q$ the
Laplacian matrix, as will be defined in section \ref{sec_defs}. In words, we
model each component by a differential equation, such that the equations of the
whole system become coupled due to the linking of the components in the
network. Assuming such a perspective, the synchronization of a network maps
into the dynamics of (\ref{eq_synch0}). The paper has been mostly written with this
type of application in mind. Nevertheless, results extend to all phenomena
dominated by the Laplacian, such as diffusion delivery of any commodity on a
network.

It is well-known that the synchronization of diffusively-coupled systems on
networks is crucially affected by the network topology \cite{Olfati-Saber2005,
Olfati-Saber2007, Simpson-Porco2012, Yamamoto2011}. However, current research
methods focus almost exclusively on individual networks treated as isolated
systems. In reality, an individual network is often a combined system of
multiple networks with distinct topologies and functions. This motivates us to
study the effect of interdependent topologies on the mutual synchronization of
networks. Recently, effort has been directed to these complex systems composed
of many interdependent networks, which seem to model complex systems better
than single networks \cite{Wang2013, Shlomo10}. For instance, a pathogen
spreads on a network of human contacts supported by global and regional
transportation networks; or in a power grid and a communication network, that
are coupled together \cite{Simpson-Porco2012}, a power station depends on a
communication node for control, while a communication node depends on a power
station for electricity. Cascading failures on interdependent networks, where
the failure of a node at one end of an interdependent link implies the failure
of the node at the other end of the link, have been widely studied
\cite{Shlomo10,Leicht2009}. The latter studies show that results obtained in
the context of a single isolated network can change dramatically once
interactions with other networks are incorporated.

In particular, we will focus on the the so-called \textit{algebraic
connectivity} of interdependent networks, which is defined as the second
smallest eigenvalue $\mu_{N-1}$ of the discrete Laplacian matrix. This
eigenvalue plays an important role on, among others, synchronization dynamics,
network robustness, consensus problems, flocking and swarming, belief
propagation, synchronization of coupled oscillators, graph partitioning,
distributed filtering in sensor networks \cite{Olfati-Saber2005,
Olfati-Saber2007, Simpson-Porco2012, Yamamoto2011, Shang2012, Olfati-Saber2005,
Freeman2006, Jamakovic2007, Fiedler1975, Strogatz2001}. In the present work, we
interpret the algebraic connectivity as the inverse of a ``proper time'', since
the deviations from equilibrium in \eqref{eq_synch0} decay exponentially with
such scale. Larger values of $\mu_{N-1}$ enable synchronization in both
discrete and continuous-time systems, even in the presence of transmission
delays \cite{Lin2008303,Shang2012}. From a graph theoretic perspective, we
will show that the algebraic connectivity experiences a phase transition upon
the addition of a sufficient number of links among two interdependent
networks. In other words, system synchronizability does not experience any
depletion when the operability of the control channel is softly reduced.

This paper is structured as follows. Section \ref{sec_defs} introduces the
necessary notation and exposes both the Laplacian matrix and the graph spectra.
Sections~\ref{sec_exactResults} and \ref{sec_perturbationTh} provide a
mean-field approach for the algebraic connectivity, and exploit the
perturbation theory of interdependent networks, respectively. Finally,
numerical results are presented in Section~\ref{sec_simulations}.
The latter will expose properties of regular, random, small-world, and
scale-free networks. Conclusions are drawn in Section~\ref{sec_concl}.

\section{Definitions} \label{sec_defs}

\subsection{Graph Theory Basics}

A graph $G$ is composed by a set of nodes interconnected by a set of links
$G\left(\mathcal{N},\mathcal{L}\right)$.
Suppose one has two networks $G_1  =  (\mathcal{N}_{1},\mathcal{L}_{1})$ and $
G_2  =  (\mathcal{N}_{2},\mathcal{L}_{2})$, each with a set of nodes
$\left(\mathcal{N}_{1},\mathcal{N}_{2}\right)$ and a set of links
$\left(\mathcal{L}_{1}, \mathcal{L}_{2}\right)$ respectively. For simplicity,
in this paper we only study the case where $G_1$ and $G_2$ are identical, i.e.
$G_1 = G_2$, meaning that the $i$-th node of $G_1$ is topologically equivalent
to the $i$-th node in $G_2$. In the following, we will suppose any dependence
relation to be symmetric, i.e. all networks are undirected.
 
The global system resulting from the connection of the two networks is a
network $G$ with $\mathcal{N}_{1} \cup \mathcal{N}_{2}$ nodes and
$\mathcal{L}_{1} \cup \mathcal{L}_{2}$ ''intralinks'' plus a number of
''interlinks'' $\mathcal{L}_{12}$ joining the two networks; that is
$\mathcal{N}=\mathcal{N}_{1} \cup \mathcal{N}_{2}$ and
$\mathcal{L}=\mathcal{L}_{1} \cup \mathcal{L}_{2}\cup \mathcal{L}_{12}$, thus
$(\mathcal{N},\mathcal{L})=G \stackrel{def}{=} (\mathcal{N}_{1} \cup
\mathcal{N}_{2}, \mathcal{L}_{1} \cup \mathcal{L}_{2}\cup \mathcal{L}_{12})$.

Let us denote $N_{i}$ as the number of nodes in $|\mathcal{N}_i|$, and $L_i$ as
the number of links in as $|\mathcal{L}_{i}|$, also $N=N_1+N_2$, and
$L=L_1+L_2$; let $A_1$ and $A_2$ be the adjacency matrices of the two networks
$G_1$ and $G_2$, and $A$ that of the whole system $G$, whose entries or
elements are $a_{ij}=1$ if node $i$ is connected to node $j$, otherwise
$a_{ij}=0$. When the two networks are disconnected ($\mathcal{L}_{12} =
\emptyset$), the matrix $A$ is defined as the $N\times N$ matrix:

\begin{equation*}
	A=\left[ \begin{array}{cc}
	A_{1} & \mathbf{0} \\
	\mathbf{0} & A_{2}%
\end{array} %
\right]. \end{equation*}

When an interaction is introduced ($\mathcal{L}_{12} \ne \emptyset$), the
adjacency matrix acquires non-trivial  off-block terms denoted by $B_{ij}$,
defined as the $N_i\times N_j$ interconnection matrix representing the
interlinks between $G_{1}$ and $G_{2}$. The interdependency matrix $B$ is then

\begin{equation*}
B=\left[
\begin{array}{cc}
	\mathbf{0} & B_{12} \\
	B_{12}^{T} & \mathbf{0}
\end{array}
\right].
\end{equation*}

When the two networks $G_1$ and $G_2$ are equal, the adjacency matrix of the
total system can be written as:

\begin{equation}
	A+\alpha B=\left[
	\begin{array}{cc}
	A_{1} & \alpha B_{12} \\
	\alpha B^T_{12} & A_{2}
	\end{array}
\right].
\label{eq_Azero}
\end{equation}

\noindent where $\alpha$ represents coupling strength of the interaction. If
the type of relation inside the $A$ and $B$ networks is the same (i.e. if they
represent infrastructures of the same type, such as for instance electric
systems), one may study the properties of non-weighted adjacency matrices.
 
Similarly to the adjacency matrix, one may introduce the Laplacian matrix
$Q=D-A$; where $D$ is the diagonal matrix of the degrees, where the degree of
the $i$-th node is \newline $d_{i}\stackrel{def}{=} \sum_{j}a_{ij}$. In the
same vein, one may define the diagonal matrices:
 
\begin{equation*}
\left\{
	\begin{array}{lll}
	(D_1)_{ii} & \stackrel{def}{=} &   \sum_{j}(B_{12})_{ij},\\
 	(D_2)_{ii} & \stackrel{def}{=} &  
	\sum_{j}(B_{21})_{ij}=\sum_{j}(B^T_{12})_{ij}; \end{array} .
\right.
\end{equation*}

\noindent and the Laplacian $Q$ of the total system $G$ reads:

\begin{equation}
Q=Q_A + \alpha Q_B=\left[
\begin{array}{cc}
	Q_{1} +  \alpha D_1& - \alpha B_{12} \\
	- \alpha B^T_{12} & Q_{2}+ \alpha D_2
\end{array}
\right].
\label{eq_LaplacianMatrix}
\end{equation}

\noindent where $Q_1=Q_2$ is the Laplacian matrix of $A_1=A_2$, and $Q_B$ is
the Laplacian only representing the interlinks:

\begin{equation}
 Q_B= D - B = \left[
\begin{array}{cc}
 D_1 &  -  B_{12} \\
 - B^T_{12} &  D_2
\end{array}
\right].
\end{equation}

\subsection{Fiedler Partitioning}

Since $Q$ is a real symmetric matrix, it has $N$ real eigenvalues
\cite{VanMieghemSpectra}, which we order non-decreasingly $0={\mu _N} \le {\mu
_{N - 1}} \le  \cdots  \le {\mu _1}$. The eigenvector $x_{N-1}$ corresponding
to the first non-zero eigenvalue $\mu_{N-1}$ provides a graph partition named
after Fiedler, who derived the majority of its properties
\cite{Fiedler1973a,Fiedler1975}. The $N-1$ largest Laplacian eigenvectors and
eigenvalues satisfy the following equations:
 
\begin{equation}
\left\{
	\begin{array}{rl}
	Q x &=\mu x, \\
	x^T x &=1,\\
	x^T u	&=0.
	\end{array}
\right.
\label{eq_laplacianEigenv}
\end{equation}

\noindent where $u$ is the all ones vector, which is the Laplacian eigenvector
belonging to $\mu_N=0$. The algebraic connectivity $\mu_{N-1}$ is the smallest
of the $N-1$ eigenvalues satisfying the equations in
\eqref{eq_laplacianEigenv}. Equivalently, $x_{N-1}$ and $\mu_{N-1}$ optimize
the quadratic form $x^T Q x$ subject to two constraints:

\begin{equation}
	\mu_{N-1}=\min_{x^2=1,x^T u=0}{x^{T} Q x}.
\end{equation}

\noindent Since we will only deal with the Fiedler eigenvector, we will
simplify the notation of the eigenpair $\left(\mu_{N-1},x_{N-1}\right)$ by
simply writing $\left(\mu,x\right)$.

\section{Exact results for mean-field theory} \label{sec_exactResults}

\subsection{Diagonal interlinking}

Let us start with the case of two exactly identical networks connected by
$\mathcal{L}_{12}$ corresponding interlinks. The mean-field approach to such a
system consists in studying a graph of two identical networks interacting via
$N_{1}$ weighted connections among all corresponding nodes. The weight of each
link, represented by $\alpha = \frac{L_{12}}{N_{1}}$, equals the fraction of
nodes linked to their corresponding in the exact network. In other words
$B_{12}= I$, such that the synchronization interdependence is modulated by the
parameter $\alpha$:

\begin{equation}
Q_B= \left[
\begin{array}{cc}
I & - I \\
- I & I
\end{array}
\right].
\end{equation}

\noindent and

\begin{equation}
Q_A+ \alpha Q_B=
\left[
\begin{array}{cc}
Q_1 +\alpha I & -\alpha I \\
-\alpha I & Q_2 +\alpha I
\end{array}
\right].
\end{equation}

In the language of physics, $ \alpha$ represents the coupling constant of the
interaction between the networks. Consistently with the rest of the paper, this
system will also be referred to as the mean-field model of the \textit{diagonal
interlinking} strategy. Regardless of its origin, this system exhibits some
interesting properties worth discussing.
 
Let $\xi_{N_1}, \xi_{N_1 - 1}, ..., \xi_{1}$ be the set of eigenvectors for the
Laplacian of the single network $A_1$, and $\omega_{N_1}, \omega_{N_1 - 1},
..., \omega_{1}$ be their relative eigenvalues. Since the perturbation $Q_B$
commutes with $Q_A$, all the eigenvectors of the interdependent graph are kept
unchanged \cite{VanMieghemSpectra}.  All the (unperturbed) eigenvalues are
degenerate in pairs and, hence, one may define a set of eigenvectors based on
those of the single networks:
 
\begin{equation}
\left\{
\begin{array}{l}
x_{2i}=
\left[
\begin{array}{c}
\xi_i \\
 \xi_i.
\end{array}
\right] \\
x_{2i+1}=
\left[
\begin{array}{c}
\xi_i \\
- \xi_i.
\end{array}
\right]
\end{array}
\right. 
\label{eigen0}
\end{equation}

\noindent The eigenvalues for the total non-interacting system (i.e. $\alpha =
0$) are the same as for the unperturbed system $\mu_{2i}=\mu_{2i+1}=\omega_i$,
hence, the ascending sequence of eigenvalues for the non-interactive system is
$\omega_{N_1}=0, 0, \omega_{N_1 - 1}, \omega_{N_1 - 1} \ldots , \omega_{1},
\omega_{1}$. When the interaction is switched on (i.e. $\alpha \ne 0$), the
even eigenvalues are kept unaltered, while the odd ones increase linearly with
$2\alpha$,

\begin{equation}
\left\{
\begin{array}{l}
\mu_{2i}= \omega_i ,\\
\mu_{2i+1}=\omega_i + 2 \alpha .
\end{array}
\right.
\end{equation}

For $\alpha$ close to zero, the eigenvector ranking is kept unchanged $\mu_N
=\omega_{N_1}=0, \mu_{N-1}=\alpha, \mu_{N-2}=\omega_{{N_1}-1}, \mu_{N-3} =
\omega_{{N_1}-1}+ 2\alpha, \ldots , \omega_{1}, \omega_{1}+ 2\alpha$.
However, when $\alpha > \frac{\omega_{{N_1}-1}}{2}$ the second and third
eigenvalues of the interdependent network ($\mu_{N-1}$ and $\mu_{N-2}$) swap.
Therefore, the first non-zero eigenvalue increases linearly with $2\alpha$ up
to the value of the isolated networks $\omega_{{N_1}-1}$ at which it reaches a
plateau.  In other words, when $\alpha$ is greater than the threshold
$\alpha_{I} = \frac{\omega_{{N_1}-1}}{2}$ the interactive system is capable of
synchronizing with the same swiftness as the single isolated network. Thus
when the system intercommunication channel is quicker than the proper time
(the inverse of the algebraic connectivity), then the proper time of the
interactive system equals that of the single network. The critical value $\alpha_{I}$ for the
exact model corresponds to a critical value of links $l_{I}$ to be included to
achieve the swiftness of the single network:

\begin{equation}
l_{I}=\alpha_{I}N_1 = \frac{\omega_{N-1} \cdot N_{1}}{2}.
\label{eq_diagonalTh}
\end{equation}

If we interpret network robustness as the ability of a system to perform its
function upon damage or attacks, then it is worth discussing what happens when
two networks, $A_1$ and $A_2$, originally fully connected by diagonal
interlinking $B$, are subject to some interlink loss. Our simple, exact model
shows that when these two fully connected networks are subject to minor
interlink loss, the response of the total interacting system $A+\alpha B$ takes
place at the same speed as the single component network $A_1$.  In other words,
when the operability of the control channel via $\alpha$ is mildly reduced,
the global system synchronizability does not decrease. However if the
operability of the connection devices degrades below the critical value
$\alpha_{I}$, the synchronization process starts to slow down. From the
mean-field approach point of view, this means that the system may lose a
fraction of interlinks while keeping its synchronization time unchanged.

Following the statistical variant, the parameter $\alpha$ can be regarded as a
coupling constant or inverse temperature. If one identifies the Fiedler
eigenvalue $\mu_{N-1}$ with the internal energy of a thermodynamical system,
then its first derivative exhibits a jump from zero to a finite value.
Nevertheless, this derivative does not diverge as expected for a second order
transition \cite{Blundell2010} \footnote{the order of a phase transition is
the order of the lowest differential which shows a discontinuity.}.

On the other hand, if one employs the Fiedler eigenvalue as a metric for the
synchronizability and regards it as a thermodynamical potential such as the
free enthalpy, its Legendre transform corresponds to the internal energy and
exhibits a discontinuity at $\alpha = \alpha_{I}$. In this perspective, one
may interpret the observed abrupt change as a first order phase transition.
Despite this interesting parallel, it is worth noting that the Fiedler
eigenvalue and its Legendre transform are not extensive quantities and, hence,
they cannot be properly regarded as thermodynamical potentials. However, the
behavior of the system closely resembles a phase transition.

To understand the intimate nature of the phase transition, one may inspect the
topological properties of the eigenvectors. Below the critical value
$\alpha_{I}$, the cut links associated to the Fiedler partition lay outside the
originally isolated networks (i.e.interlinks are cut), whereas just above the
critical value, all cut links lay inside the originally isolated networks (i.e.
intralinks are cut). This means that, below $\alpha_{I}$, the synchronization
is dominated by the intralinks in $\alpha B$, while beyond $\alpha_{I}$ the
synchronization involves the whole system, $A+\alpha B$.

\begin{figure*}[t]
\centering
\includegraphics[width=0.95\textwidth]{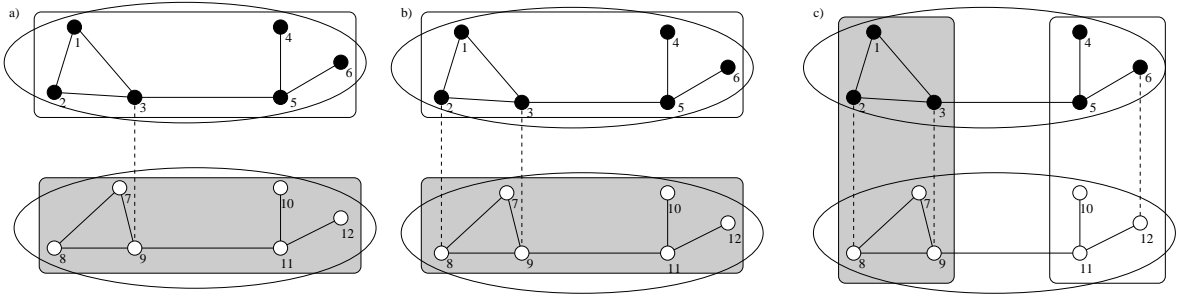}
\caption{Two graphs with $6$ black and white nodes respectively and $7$ links
each are progressively interconnected with \textit{a)} $1$ interlink,
\textit{b)} $2$ interlinks and \textit{c)} $3$ interlinks. Adding $1$ or $2$
interlinks causes the Fiedler eigenvector to split (depicted by the rectangles)
the network into the natural partitions $G_1$ and $G_2$ . For both cases, the
confining links match the added interlinks (dashed lines). However, when adding
$3$ interlinks the Fiedler partition experiences a brusque shift, causing the
intralinks of the single networks to become the confining links. Thus the added
interlinks become a part of the Fiedler partitions.}
\label{fig_manual_display}
\end{figure*}

\subsection{General interlinking}

A second important example that may be treated algebraically corresponds to the
mean-field approximation of the \textit{general interlinking} strategy. The
mean-field approach consists of studying a graph with two identical networks
interacting via $N_1^2$ weighted connections. The interdependence matrix is a
matrix with all unitary components: $B_{12}=J$, where $J$ is the all ones
matrix; the weight of each interlink is $\alpha=\frac{L_{12}}{N_1^2}$, and
 
\begin{equation}
Q=Q_A+\alpha Q_B=
\left[
\begin{array}{cc}
Q_1 +\alpha N_1 I & -\alpha J \\
-\alpha J & Q_2 +\alpha N_1 I
\end{array}
\right].
\label{eq_generalCase}
\end{equation}

As in the previous case, the $Q_B$ matrix commutes with $Q_A$ and hence a
common set of eigenvectors can be chosen as in (\ref{eigen0}). The null
eigenvalue $\mu_N$ is always present, while all the others experience some
increase for a non-trivial $\alpha$: all eigenvalues $\mu_i$ for $i$ smaller
than $N-1$, increase for a fixed amount $\alpha N_1 $, while $\mu_{N-1}$
increases by twice that quantity,
 
\begin{equation}
\left\{
 \begin{array}{ll}
\mu_{N} & =0,  \\
\mu_{N-1} &= 2\alpha N_1,  \\
\mu_{i} &=\omega_i + \alpha N_1 \text{, for $i \le N-1$. }
\end{array}
\right.
\end{equation}

This different rate of growth again implies that there exists a critical value
$\alpha_{J}$ beyond which the second and third eigenvectors ($\mu_{N-1}$ and
$\mu_{N-2}$) swap. The threshold $\alpha_{J}$ can be easily calculated imposing
the crossing condition $\mu_{N-1}=\mu_{N-2}$:
 
\begin{equation}
\alpha_{J}=\frac{\omega_{N-1}}{N_1}.
\end{equation}

With $\alpha=\frac{L_{12}}{N_1^2}$, the critical number of links for the
general interlinking strategy can be also estimated in the mean-field
approximation:
 
\begin{equation}
l_{J}=\alpha_J N_1^2 =\omega_{N-1} \cdot N_1.
\label{eq_generalTh}
\end{equation}

It is worth noting that, the critical number of interlinks corresponding to the
mean-field theory of the diagonal (\ref{eq_diagonalTh}), and general
(\ref{eq_generalTh}) interlink strategies, differ simply by a factor of $2$.
\section{Approximating $\mu_{N-1}$ using perturbation theory}
\label{sec_perturbationTh}
 
The problem consists in finding the minimum of the associated quadratic form in
the unitary sphere ($x^Tx=1$), with the constraint $u^T x = 0$.

\begin{equation}
	\mu=\mu_{N-1} = \inf_{x \ne 0, u^T x=0}\frac{x^{T}Qx}{x^{T}x};
\label{eq_Lspectra}
\end{equation}

In our case, the matrix $Q$ is the sum of a matrix $Q_A$ linking only nodes
inside the same net, and a ``perturbation'' $\alpha Q_B$ that only connects
nodes in different networks ($Q_A+\alpha Q_B$). Therefore, we want to find the
minimum that satisfies the spectral equations:
 
\begin{equation}
\left\{
\begin{array}{rcr}
(Q_A+ \alpha Q_B -\mu I)x &=0, \\
x^{T} x &=1, \\
u^T x &=0.
\end{array}
\right.
\end{equation}

\noindent When the solution is analytical in $\alpha $, one may express $\mu$
and $x$ by Taylor expansion as

\begin{eqnarray}
\mu &=&\sum_{k=0}^{\infty }\mu^{(k)}\alpha ^{k} \\
x &=&\sum_{k=0}^{\infty }x^{(k)}\alpha ^{k}
\end{eqnarray}

\noindent Substituting the expansion in the eigenvalue equation
(\ref{eq_Lspectra}) gives the hierarchy of equations:

\begin{equation}
\left\{
\begin{array}{lccl}
Q_A x^{(k)}+\alpha Q_Bx^{(k-1)}& = &\sum_{i=0}^{k}\mu ^{(k-i)}x^{(i)}&
\textit{for all k},\\
\sum_{i=0}^{k}x^{(k-i)}x^{(i)}& = &0 & \textit{for  $k \ge 1$}, \\
u^T x^{(k)} &=& 0 & \textit{for all k}.
\end{array}
\right.
\label{eq_hierarchy}
\end{equation}

\subsection{Explicit approximations up to the second order}

The zero order expansion just provides a simple set of equations:

\begin{equation}
\left\lbrace
\begin{array}{lll}
Q_A  x^{(0)} &= \mu ^{(0)}  x^{(0)}, \\
 x^{(0)}  x^{(0)} &=1, \\
u^T  x^{(0)} &=0.
\end{array}
\right.
\label{zeroth}
\end{equation}

Let $\left(\mu_{N-1}\right)_{A1}, \left(\mu_{N-1}\right)_{A2}$ and
$\left(x_{N-1}\right)_{A1}, \left(x_{N-1}\right)_{A2}$ denote the smallest
non-zero eigenvalue and the corresponding eigenvector of $Q_{1}, Q_{2}$,
respectively. Similarly
 
\begin{equation}
\left\{
\begin{array}{lcl}
(x_{N_1})_{A1}&=&1/\sqrt{N_1}(1,1,\ldots ,1,0,0,\ldots ,0),\\
(x_{N_2})_{A2}&=&1/\sqrt{N_2}(0,0,\ldots ,0,1,1,\ldots ,1).
\end{array}
\right.
\end{equation}

\noindent will represent the null eigenvectors of network $G_1$ and $G_2$,
respectively. When the networks are put together, any combination of the former
is a null eigenvector. Two special combinations are worth employing: the
trivial solution corresponding to the constant vector:

\begin{equation}
x_{N}=\frac{1}{\sqrt{N}}(1,\ldots, 1)=
\sqrt{\frac{N_1}{N}}(x_{N_1})_{A1}+\sqrt{\frac{N_2}{N}}(x_{N_2})_{A2}.
\label{eq_eigenVERSOR}
\end{equation}

and the other combination orthogonal to the former that represents a useful
starting point for the perturbation theory:

\begin{equation}
x_{N-1}^{(0)} = x^{(0)} =\frac{1}{\sqrt{N}}(1,\ldots ,1,-1,\ldots ,-1)=
\sqrt{\frac{N_1}{N}}(x_{N_1})_{A1}-\sqrt{\frac{N_2}{N}}(x_{N_2})_{A2}.
\end{equation}

\noindent which satisfies the zero order approximation (\ref{zeroth}).
The zero order approximation to the Fiedler eigenvalue is then null:

\begin{equation}
 \mu ^{(0)}=0.
\end{equation}

The first order approximation equations follow from (\ref{eq_hierarchy}) as:

\begin{equation}
\left\lbrace
\begin{array}{rcl}
        Q_A x^{(1)}+\alpha Q_{B}x^{(0)}&=& \mu^{(1)}x^{(0)} \\
	\left( x^{(0)}\right)^T x^{(1)} &=&0\\
	u^{T} x^{(1)} &=&0.
\end{array}
\right.
\label{eq_firstOrderEquations}
\end{equation}

Taking the projection over $x^{(0)}$ of the first equation of
(\ref{eq_firstOrderEquations}), one obtains the first order correction
$\mu^{(1)}$ that depends on the zero order eigenvector only:
 
\begin{equation}
 \mu^{(1)} =\left( x^{(0)}\right)^T \alpha Q_B x^{(0)}
\end{equation}

A simple case to analyze is that where only one interlink joins $A_1$ with
$A_2$: $(B_{12})_{ij}=\delta_{ik}\delta_{kj}$; in this case
$(d_1)_{kk}=\delta_{ik}$ and $(d_2)_{ll}=\delta_{jl}$ and the perturbation
estimate gives:
 
\begin{equation}
	\mu^{(1)} =(\frac{1}{2}(1+1)+1) (\eta_i)^2=\frac{2}{N_1} \geq \mu_{N-1}(Q).
\label{eq_1linkFirstOrderApprox}
\end{equation}

\noindent where $\eta$ is the single net ($N_1$ dimensional) unitary vector
$\eta\stackrel{def}{=} 1/\sqrt{N_1}(1,1,\ldots ,1)$. When $k$ interlinks are
included, $Q_B$ is just the sum of $k$ contributions of the previous type thus
$\mu^{(1)} = \frac{2k}{N_1}$. That is, the first order correction to the
Fiedler eigenvalue increases linearly with the number of interlinks.
The first order correction to the eigenvector can be evaluated from
(\ref{eq_firstOrderEquations}) as a solution of the linear equation:
 
\begin{equation}
	Q_A x^{(1)}=-\left(\alpha Q_{B}-\mu^{(1)}\right)x^{(0)}.
\label{eq_firstOrderX}
\end{equation}

\noindent where the operator $Q_A$ is invertible out of its kernel ($Q_A v
=0$); since $ \left(\alpha Q_{B}-\mu^{(1)}\right)x^{(0)} $ is orthogonal to the
kernel, (\ref{eq_firstOrderX}) is solvable.

The second order equations follow from (\ref{eq_hierarchy}) as

\begin{equation}
\left\lbrace
\begin{array}{rl}
	Q_A x^{(2)}+  \alpha Q_B x^{(1)}  &=\mu ^{(0)} x^{(2)}+ \mu^{(1)}x^{(1)}+\mu
	^{(2)}x^{(0)} \\
	\left( x^{(0)} \right)^T x^{(2)}+  \left(x^{(1)}\right)^Tx^{(1)}
	+\left(x^{(2)}\right)^T x^{(0)} &=0\\
	u x^{(2)} &=0
\end{array}
\right.
\end{equation}

\noindent that is, the second order correction is quadratic and equals:

\begin{equation}
	\mu ^{(2)}= \left(x^{(0)} \right)^T \alpha Q_B \left(x^{(1)}
	\right)=-\left(x^{(1)} \right)^T Q_A \left(x^{(1)} \right) \le 0.
\end{equation}

\begin{figure*}[t]
\centering
\subfloat[Diagonal strategy]{
\includegraphics[width=0.45\textwidth]{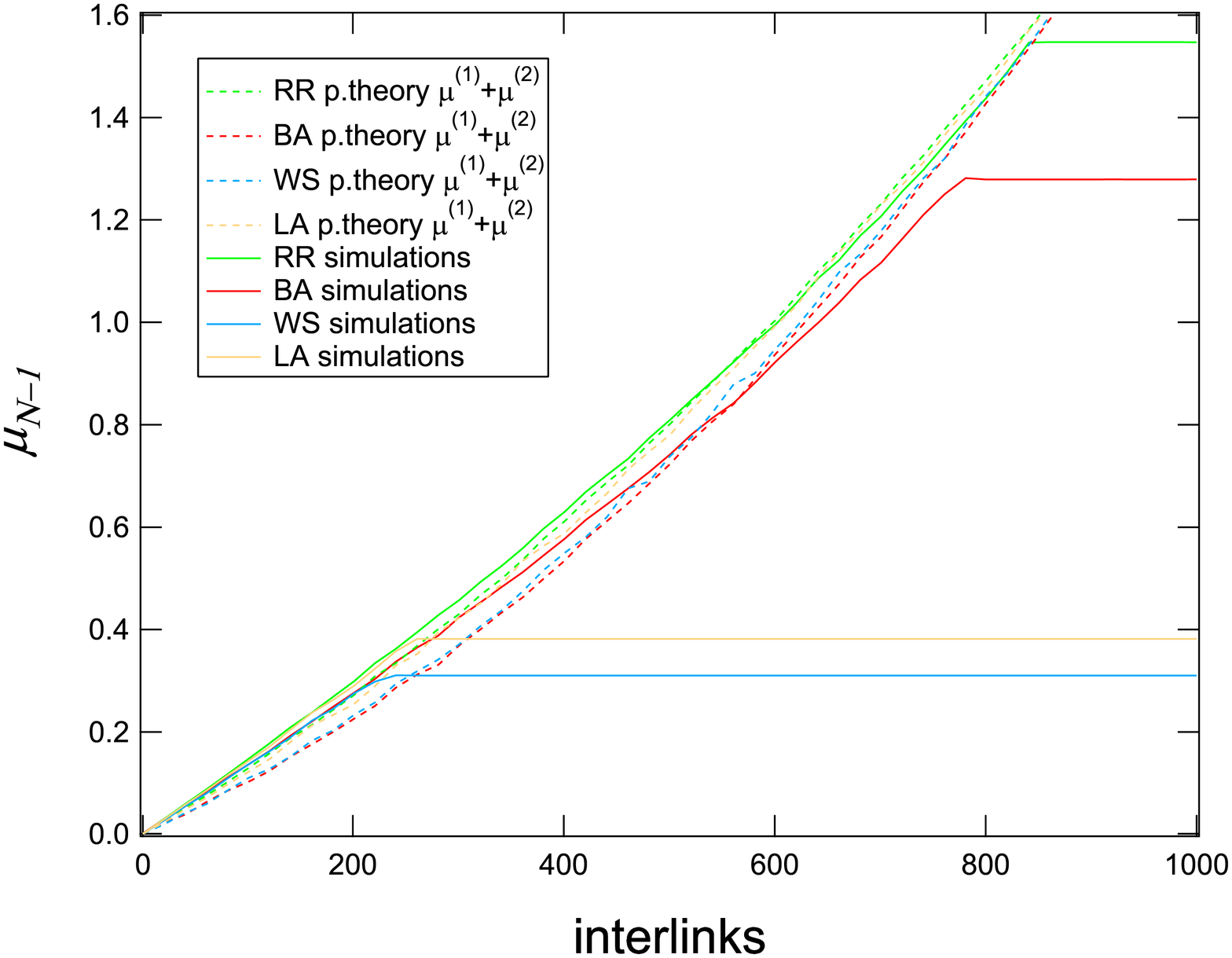}}
\subfloat[General strategy]{
\includegraphics[width=0.45\textwidth]{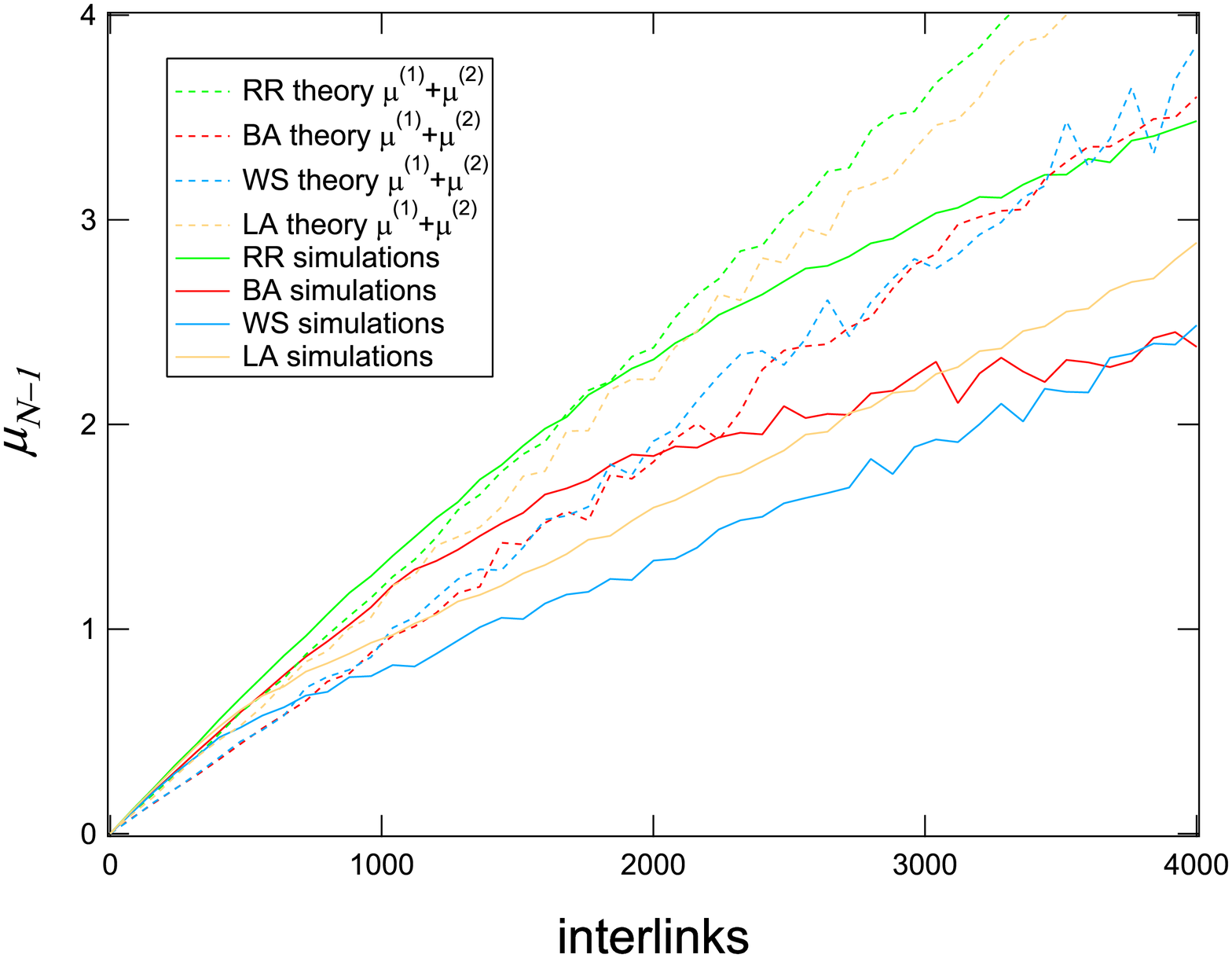}}
\caption{Simulated (solid lines) and theorized (dashed lines) algebraic
connectivity $\mu_{N-1}$ of four graph models with $N=1000$ nodes, as
interlinks are added between single networks following two strategies:
diagonal interlinks (left image) and general interlinks (right image).
Perturbation theory best approaches $\mu_{N-1}$ for the parabolic region of the
diagonal interlinks strategy, which saturates after adding $\frac{\alpha_{Th}
\cdot N}{2}$ links, as we detailed in section \ref{sec_exactResults}.}
\label{fig_theory}
\end{figure*}

\noindent As expected $\mu ^{(2)}$ is negative, thus improving the estimate of
the algebraic connectivity.  The former perturbation estimates are illustrated
in Fig.~\ref{fig_theory} together with numerical simulations.

Perturbation theory may also be applied to any initial eigenvector of
the unperturbed networks. Different perturbations $\alpha B$ will have different
effects on the quadratic form of (\ref{eq_Lspectra}) associated with all
initial eigenvectors. Therefore, it may happen that the perturbed value of
$\mu$ obtained starting from $x^{(0)}$ is smaller than the quadratic form
associated with the $x_{N-1}$ (the unperturbed eigenvector in (\ref{eigen0}))
or some other educated guess. This is precisely the origin of the phase
transition.
 
The estimates resulting form the second order perturbation theory are compared
in Fig.~\ref{fig_theory} with the results of numerical calculations. As can be
seen, for both the diagonal and the general strategies the agreement is good up
to the phase transition where the starting point of the perturbation theory
should be changed.
 
\subsection{Perturbative approximations and upper bounds}

Since we are dealing with a constraint optimization problem, finding a minimum
of a positive form, any test vector $v$ provides an upper bound for the actual
minimum value:

\begin{equation}
	\mu=\mu_{N-1} \le  \frac{v^{T}Qv}{v^{T}v}.
\end{equation}

The perturbation theory provides natural candidates as test vectors. The zero
order solution provides the simplest inequality:

\begin{equation}
	\mu_{N-1}(Q)\le  \alpha \frac{(x^{(0)})^{T}Qx^{(0)}}{(x^{(0)})^{T}x^{(0)}}=
	\alpha \mu^{(1)}.
\end{equation}

The first order approximation provides a better (i.e. lower) upper bound:

\begin{equation*}
	\mu_{N-1}(Q)\le  \frac{(x^{(0)}+\alpha x^{(1)})^{T}Q(x^{(0)}+\alpha x^{(1)})}
	{(x^{(0)}+\alpha x^{(1)})^{T}(x^{(0)}+\alpha x^{(1)})}.
\end{equation*}

that is:

\begin{equation}
	\mu_{N-1}(Q)\le  \frac{\alpha \mu^{(1)}+\alpha^2 \mu^{(2)}+ \alpha^3
	(x^{(1)})^{T}Q_Bx^{(1)}}{1+\alpha^2 (x^{(1)})^2}.
\end{equation}

which for small enough $\alpha$ is always lower than $\alpha \mu^{(1)}$.

\section{Simulations} \label{sec_simulations}

Previous sections provided basic means to understand the dependence of the
algebraic connectivity on the topology of the interdependence links. In this
section we will introduce model networks to test the predictability and the
limits of the mean-field and the perturbation approximations.

\subsection{Interdependent networks model}

Our interdependent network model consists of two main components: a network
model for the single networks, and the rules by which the two networks are
linked. In other words, to model two interdependent networks one needs to
select two model networks and one interlinking strategy.

In the numerical simulations discussed here, we considered four different graph
models for our coupled networks. These models exhibit a wide variety of
topological features and represent the four different building blocks:

\begin{itemize}
	\item \textbf{Random Regular (RR)}: random configuration model introduced by
	Bollobas \cite{Bollobas2001}. All nodes are initially assigned a fixed degree
	$d_i=k, i\in\mathcal{N}$. The $k$ degree stubs are then randomly
	interconnected while avoiding self-loops and multiple links.
	\item \textbf{Barab\'{a}si-Albert (BA)}: growth model proposed by Barab\'{a}si
	\textit{et. al.} \cite{A.L.Barabasi1999} whereby new nodes are attached to $m$
	already existing nodes in a preferential attachment fashion.
 	For large enough values of $N$, this method ensures the emergence of
 	\textit{power-law} behavior observed in many real-world networks.
	\item \textbf{Watts-Strogatz (WS)}: randomized circular lattice proposed by
	Watts \textit{et. al.} \cite{WatStr98} where all nodes start with a fixed
	degree $k$ and are connected to their $\frac{k}{2}$ immediate neighbors. In a
	second stage, all existing links are rewired with a small probability $p$,
	which produces graphs with low average hopcount yet high clustering
	coefficient, which mimics the \textit{small-world} property found in
	real-world networks.
	\item \textbf{Lattice (LA)}: a deterministic three-dimensional grid which
	loops around its boundaries (i.e. a geometrical torus).
\end{itemize}

\noindent The input parameters for each model are set such that all graphs have
the same number of nodes and links. In addition, all simulated graphs consist of
a single connected component, i.e. random graphs containing more than one
connected component were discarded.

We define two strategies to generate the interdependency matrix $B$, which we
analytically solved in section \ref{sec_exactResults}:

\begin{itemize}
	\item \textit{diagonal interlinking} strategy: links are randomly added to the
	diagonal elements of $B$, thus linking single network's analogous nodes.
	\item \textit{general interlinking} strategy: random links are added to $B$
	without restrictions, generating a random interconnection pattern.
\end{itemize} 

\noindent In the next sections, we explore the effects of the two interlinking
strategies on the Fiedler partition. However, some results can be extended to
more complicated situations, including different synthetic networks or more
complex linking strategies.

\subsection{Partition quality metrics}

Let us introduce some preliminary definitions, required for understanding of
our numerical results. We define a graph bipartition of $G$ as the two disjoint
sets of nodes $\left\{\mathcal{R},\mathcal{S}\right\}$, where $\mathcal{R} \cup
\mathcal{S} = \mathcal{N}$. We define the \textit{natural partition} of $G$ as
the partition with the two original node sets: $\mathcal{R}=\mathcal{N}_1$, and
$\mathcal{S}=\mathcal{N}_2$. The number of nodes in $\mathcal{R}$ and
$\mathcal{S}$ is counted by their cardinality $\left|\mathcal{R}\right|$ and
$\left|\mathcal{S}\right|$, respectively. In addition, we express the number of
links with one end node in $\mathcal{R}$ an another end node in $\mathcal{S}$
as
$l\left(\mathcal{R},\mathcal{S}\right)=l\left(\mathcal{S},\mathcal{R}\right)$.
Fiedler partitioning bisects the nodes in $\mathcal{N}$ into two clusters, such
that two nodes $i$ and $j$ belong to the same cluster if ${x_i}{x_j} > 0$, i.e.
the corresponding components of the Fiedler eigenvector $x$ have the same sign.
For example, if the coupling strength $\alpha$ in (\ref{eq_LaplacianMatrix}) is
zero, the bipartition resulting from Fiedler partitioning is equivalent to the
two natural clusters, i.e. $\mathcal{R}=G_1$ and $\mathcal{S}=G_2$.

\begin{figure}[t]
	\centering
	\includegraphics[width=0.45\textwidth]{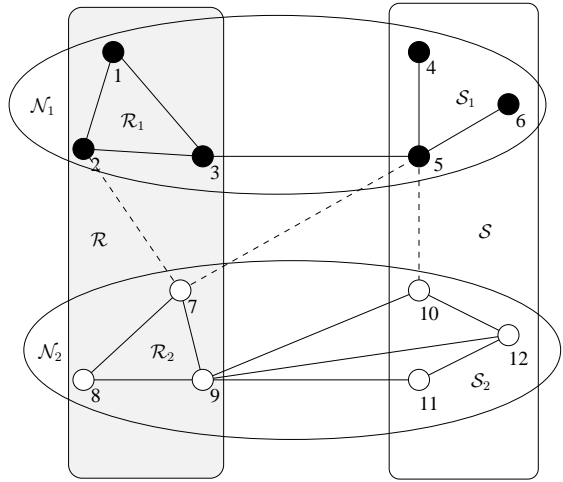}
	\caption{The four main partition sets are displayed: $\mathcal{N}_1$ (set of
	black nodes), $\mathcal{N}_2$ (set of white nodes), $\mathcal{R}$ (set of
	nodes within the gray rectangle), and $\mathcal{S}$ (set of nodes within the
	white rectangle). Both, the partition sets and the interlinks (dashed lines)
	were arbitrary chosen for illustration purposes and do not represent the
	corresponding Fiedler partition.}
	\label{fig:Illustration}
\end{figure}

The intersection between the two single graphs $\left(\mathcal{N}_1,
\mathcal{N}_2\right)$ and the Fiedler partitions $\left(\mathcal{R},
\mathcal{S}\right)$ of the interdependent network yields four node subsets,
defined as follows and illustrated in Fig.~\ref{fig:Illustration}:
\textit{(a)} $\mathcal{R}_1$ as the intersection between the positive Fielder
partition and $\mathcal{N}_1$, (b) $\mathcal{S}_1$ as the intersection between
the negative Fielder partition and $\mathcal{N}_1$,  \textit{(c)}
$\mathcal{R}_2$ as the intersection between the positive Fielder partition and
$\mathcal{N}_2$, \textit{(d)} $\mathcal{S}_2$ as the intersection between the
positive Fielder partition and $\mathcal{N}_2$. By construction the four
defined groups are disjoint, and the union of all groups equals the full set of
nodes.

In order to quantify the properties of the Fiedler partition, we study the
following set of metrics:

\begin{itemize}
	\item \textit{Fiedler cut-size} $ \stackrel{def}{=} \frac{{l\left(
	{{{\mathcal{R}}},{{\mathcal{S}}}} \right)}}{{{L_1} + {L_2}}}$.
	It represents the fraction of links with one end in $\mathcal{R}$ and another
	end in $\mathcal{S}$ (irrespective of the directionality of the link) over the
	starting number of links.
	\item \textit{interdependence angle}, defined as the angle between the
	normalized Fielder vector $x$ and the versor $x^{(0)}$, introduced in
	\eqref{eq_eigenVERSOR}. The interdependence angle is minimized when the Fiedler
	vector is parallel to the natural partition, i.e. $x^{(0)}$.
	\item \textit{entropy} of the squared Fiedler vector components $\mathop  =
	\limits^{def}  - \sum\nolimits_{i = 1}^N {{x_i}^2\log {x_i}^2} $. Based on
	Shannon's information theory metric, the entropy indicates how homogeneous the
	values in $x$ are, similarly to the participation ratio or vector localization.
	The higher the entropy, the lower is the dispersion among the values in $x$.
\end{itemize}

Some partition quality metrics may be undefined if the Laplacian matrix $Q$ is
defective \cite{Wilkinson1965}. In particular, if the second and third largest
eigenvalues of $Q_{A}+ \alpha Q_{B}$ are equal $\mu_{N-1} = \mu_{N-2}$ then any
linear combination $x' = a x_{N-1} + b x_{N-2}$ is also an eigenvector of $Q$
with eigenvalue $\mu_{N-1}$, thus the Fiedler vector is not uniquely defined.
However we will ignore these cases, which tend to occur only in graphs with
deterministic structures (e.g. the cycle graph \cite{VanMieghemSpectra}).

\subsection{Diagonal Interlinking Strategy}

\subsubsection{Strategy Description}

The \textit{diagonal interlinking} strategy consists of adding links between
the respective components of two identical networks. We can add as little as
$1$ link and as many as $N$ links.  This strategy was chosen to achieve the
maximum effect by meticulously adding a small number of interlinks. A
simplified physical example would be that of two flat metal plates: a hot one,
and a cold one. If the objective is to equalize their temperature as fast as
possible, we should adjust the plates side by side so as to maximize the heat
transfer, which is equivalent to the diagonal interlinks strategy.

\subsubsection{Initial and final states}

\begin{figure*}[!ht]
\centering
\subfloat[Algebraic Connectivity]{
\includegraphics[width=0.45\textwidth]{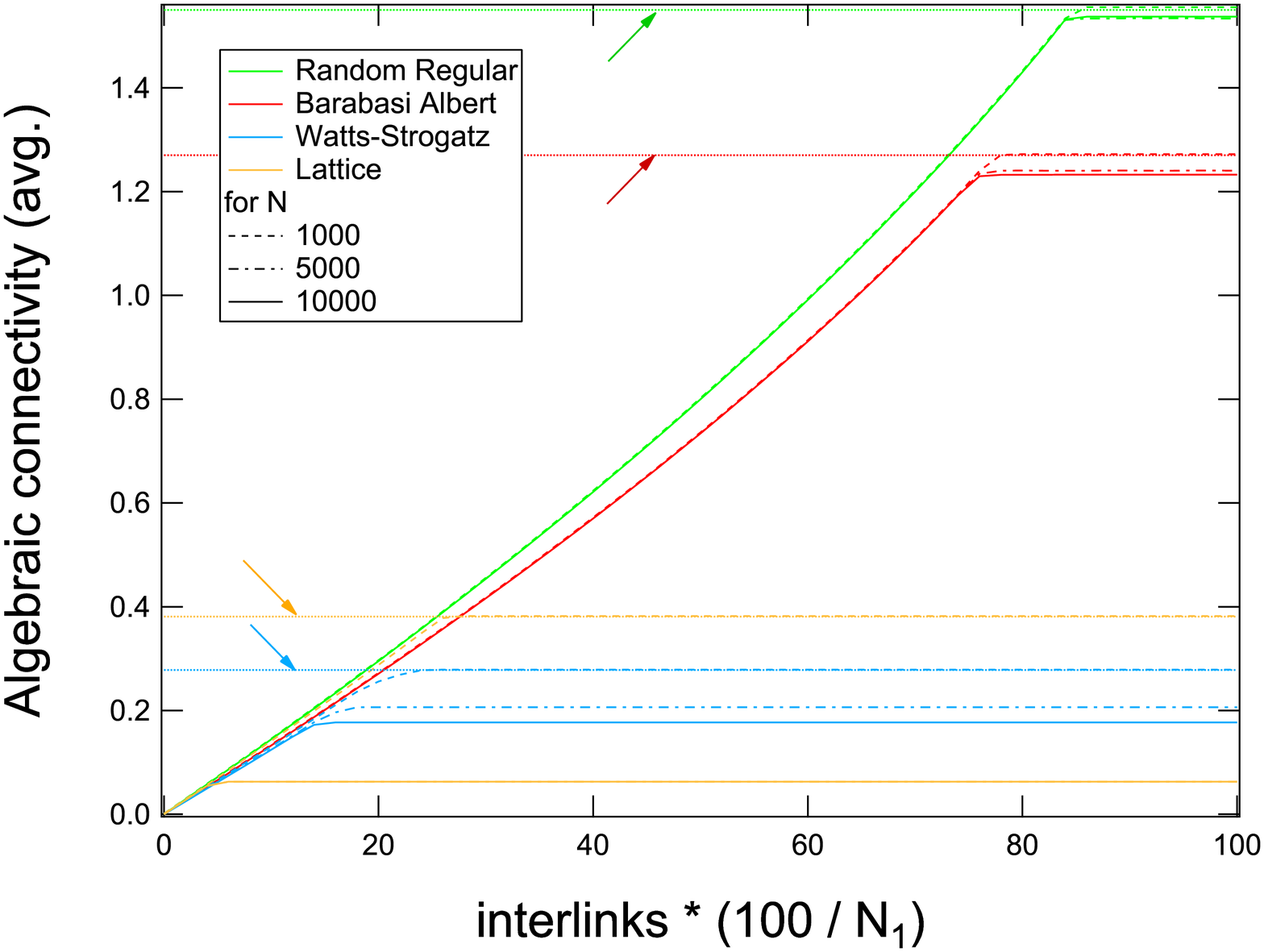}}
\subfloat[Fiedler cut]{
\includegraphics[width=0.45\textwidth]{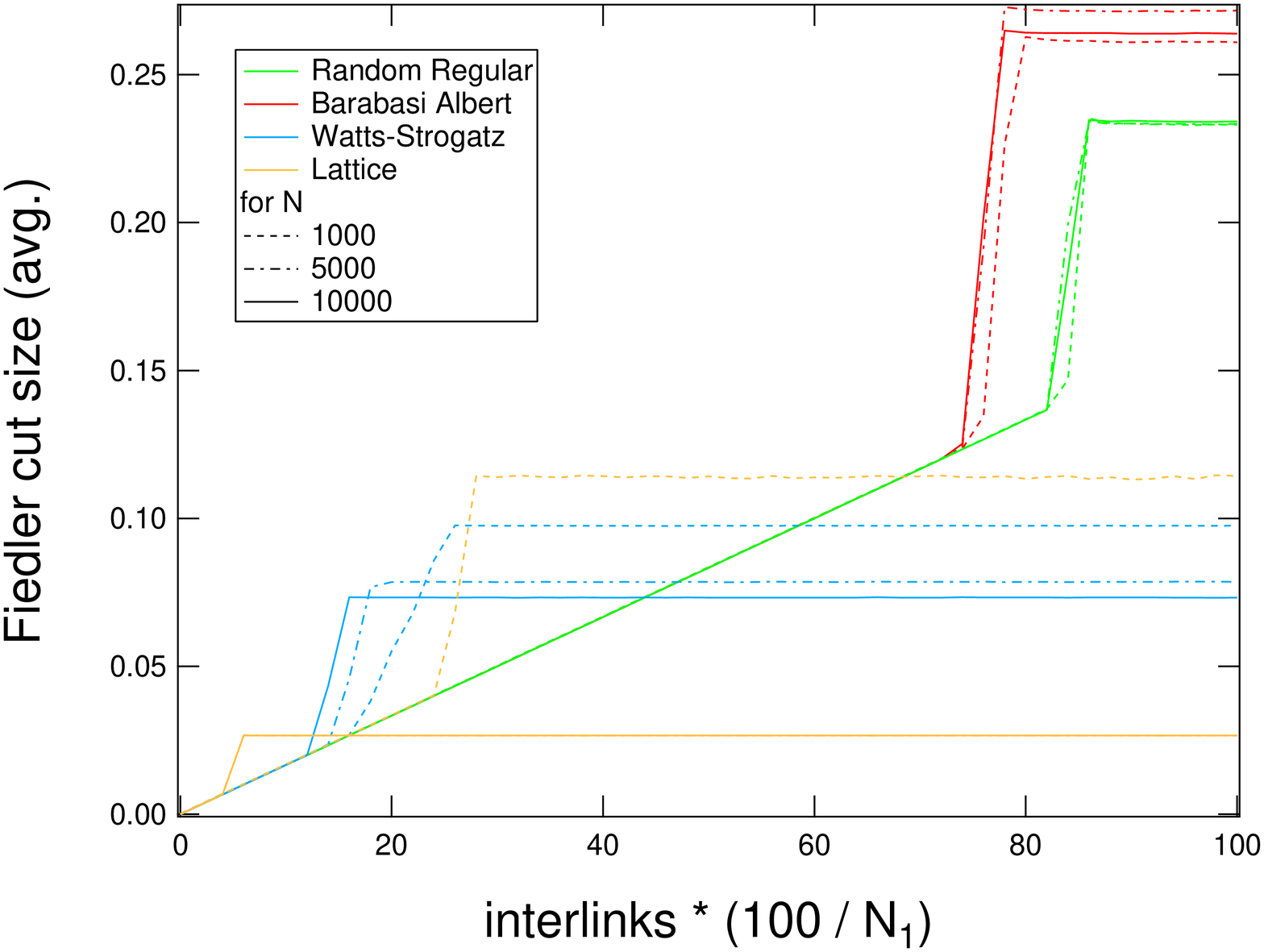}}
\\
\subfloat[Interdependence angle]{
\includegraphics[width=0.45\textwidth]{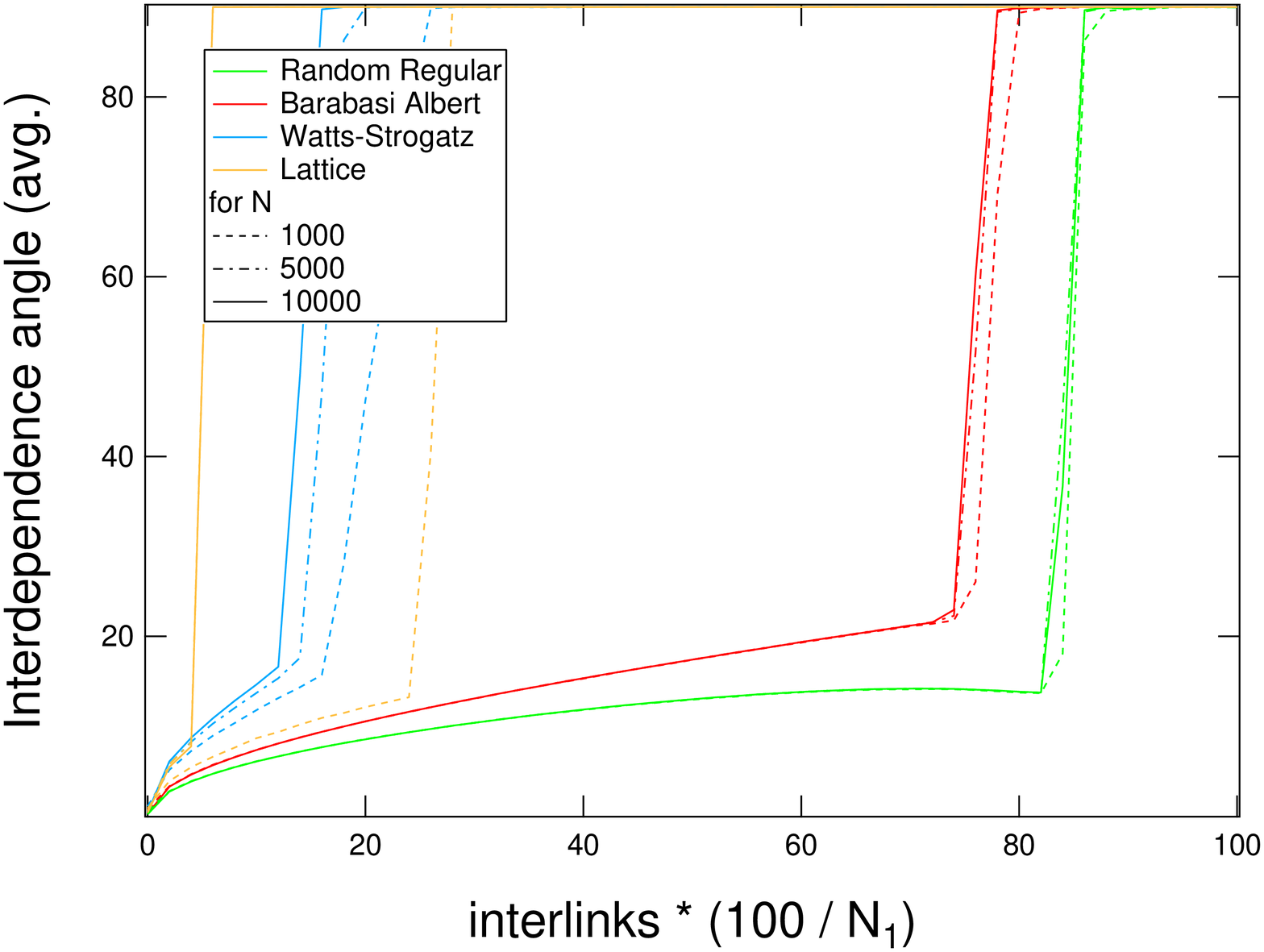}}
\subfloat[Entropy]{
\includegraphics[width=0.45\textwidth]{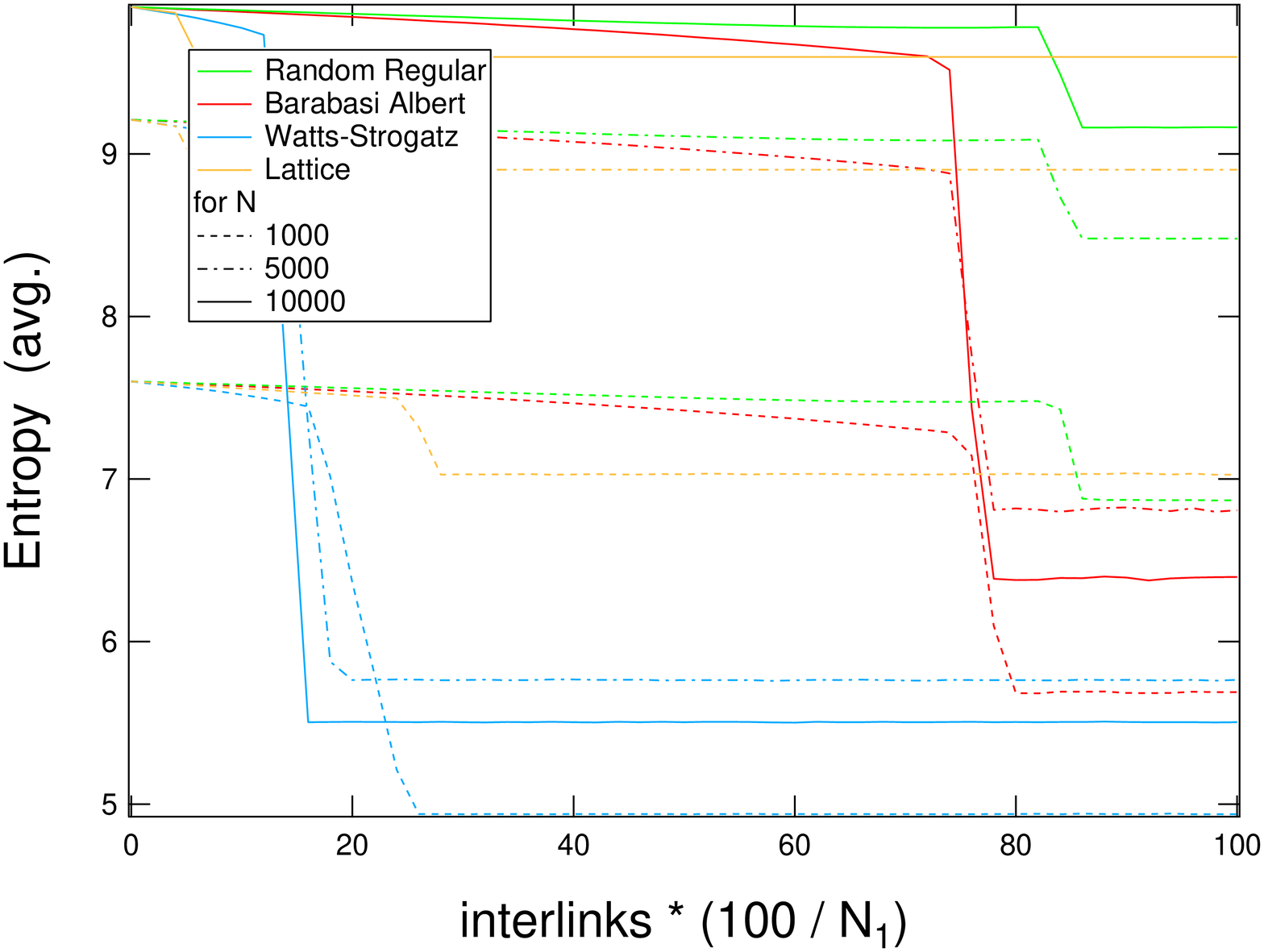}}
\caption{(Color online) Four metrics' averages are displayed to evaluate the
effect of adding interlinks following the diagonal strategy: algebraic
connectivity ($\mu_{N-1}$), Fiedler cut (${l\left(
{{{\mathcal{R}}},{{\mathcal{S}}}} \right)}/{{L_1} + {L_2}}$), interdependence
angle ($acos\left(x^T x^{(0)} / \left\| x \right\| \left\|x^{(0)}\right\|
\right)$), and entropy ($- \sum\nolimits_{i = 1}^N {{x_i}^2\log {x_i}^2}$). All
metrics experience a transition that sharpens for increasing $N$. BA and RR
graphs transition around $~80\%$ added interlinks, whereas WS and LA graphs
transition around $~20\%$. The size of the network $N_1$ has a relatively
little impact on BA and RR curves, which suggests that the transition is
independent of the network size $N_1$. The flat lines signaled with arrows in
the top left plot benchmark the average algebraic connectivity of the
$N_1=1,000$ respective single networks.}
\label{fig_diagonal_results_average}
\end{figure*}

\begin{figure*}[!ht]
\centering
\subfloat[Fiedler cut]{
\includegraphics[width=0.45\textwidth]{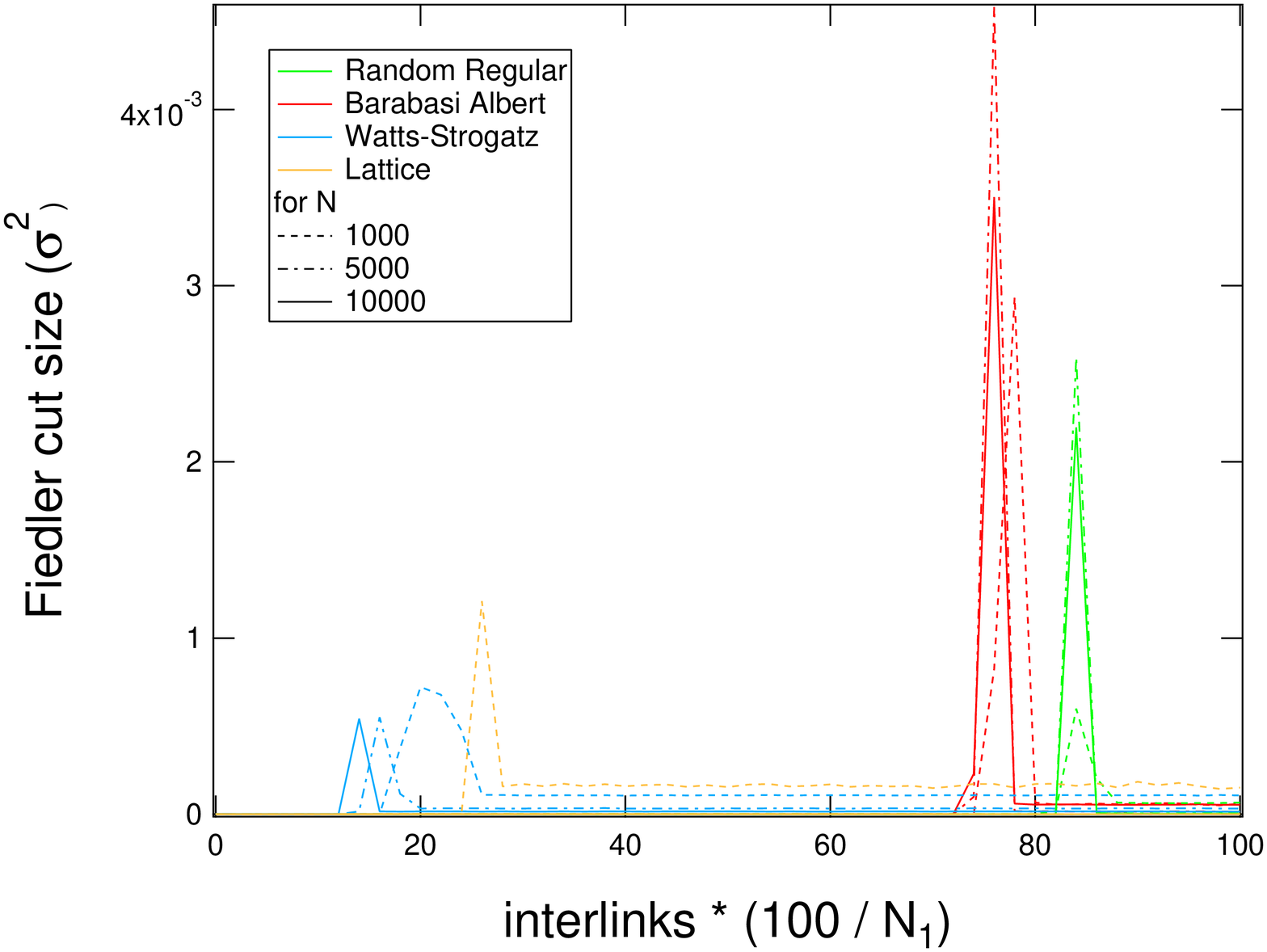}}
\subfloat[Interdependence angle]{
\includegraphics[width=0.45\textwidth]{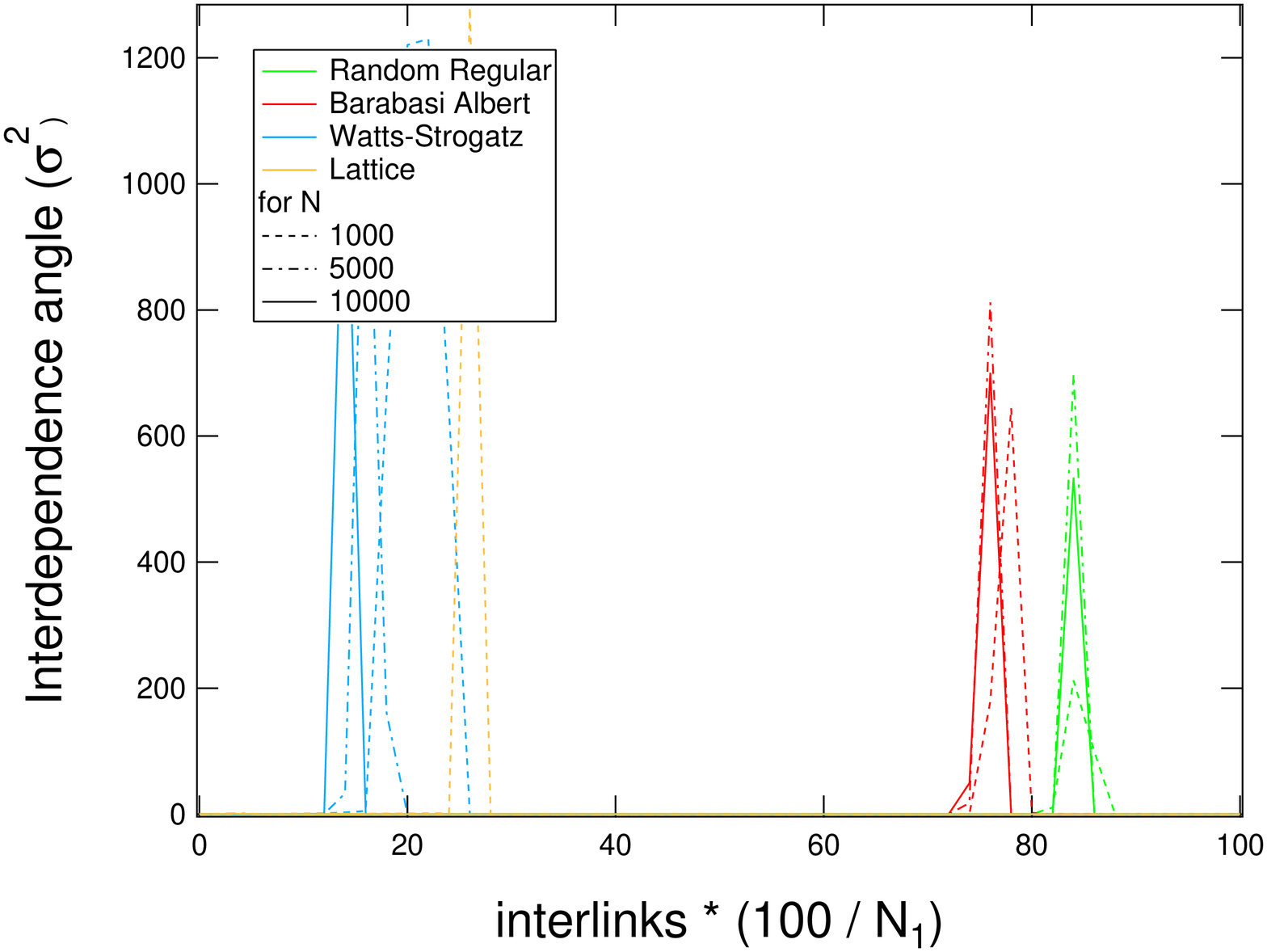}}
\caption{(Color online) The fluctuation $\sigma^2$ of the Fiedler cut and the
interdependence angle are displayed to evaluate the effect of adding interlinks
following the diagonal strategy. The narrowing peaks illustrate the sharpening
of the transition observed in Fig.~\ref{fig_diagonal_results_average}.}
\label{fig_diagonal_results_stdeviation}
\end{figure*}

We will refer to the \textit{natural},  \textit{initial} or \textit{unperturbed}
state as the scenario where there exist no interlink connecting the two networks
$G_1$ and $G_2$. The left hand side of Fig.~\ref{fig_manual_display} and
Fig.~{\ref{fig_diagonal_results_average}} illustrate the network configuration
and partition quality metrics, respectively. In this case the algebraic
connectivity dips to a null value, as no communication is possible; the Fiedler
partition then becomes undetermined. However, the sign of $x^{(0)}$ in
(\ref{eq_firstOrderX}) allows splitting the network into two clusters ${\cal P}
= G_1$ and ${\cal Q} = G_2$, corresponding to the isolated component networks.

The \textit{final} state of the diagonal interlink strategy corresponds to
$100\%$ or $N$ added interlinks, thus $B=I$ and the Fiedler vector becomes the
vector $\left(x_{N-1}(Q_1),x_{N-1}(Q_1)\right)$, as demonstrated in section
\ref{sec_exactResults}. The final partition depends exclusively on $G_1$ and
$G_2$, independently of $B$. Since we assume $G_1=G_2$, the final cut consists
purely of a subset of intralinks of $A_1$, as illustrated in the right hand
side of Fig.~\ref{fig_manual_display}. By adding (or removing) interlinks
between the two independent networks we switched from a purely interlink cut to
a purely intralink cut, which is the essence of the phase transition.

The sudden partition quality metric shifts reflect the change observed in the
Fiedler partition, while providing evidence for the existence of the
transition. This is testified by the narrowing of the region where the shifts
occurs while increasing (or decreasing) $N$ in
Fig.~\ref{fig_diagonal_results_average}. Similarly, the fluctuations of the
same quantities exhibit shrinking peaks as can be seen in
Fig.~\ref{fig_diagonal_results_stdeviation}. Upon increasing the size $N$ of
the system, the transition point seems to approach an asymptotic value. As
discussed in section \ref{sec_exactResults}, the mean-field theory predicts
this critical value to be $l_{I}=\frac{\mu_{N-1}(Q_1) \cdot N_{1}}{2}$. Since $
\mu_{N-1}(Q_1)$ has a non-trivial lower bound for increasing $N$ \cite{Wu2001},
and its algebraic connectivity is kept unchanged by the perturbation, there
will be a critical number of links beyond which $\mu_{N-1}(Q)$ does not change.
This critical point corresponds to the transition from an interlink cut to an
intralink cut.

The precise location of the jump in the simulated experiment, i.e. the critical
value of interlinks per node, depends on the graph model. However, the phase
transition is a general phenomenon, which occurrence only depends on the fact
that there exists a Fiedler cut for the single networks.

\subsubsection{Effect on partition quality metrics}

We further investigate the properties of the phase transition by looking at how
partition metrics in Fig.~\ref{fig_diagonal_results_average} evolve as
interlinks are added to the $B$ matrix.

The algebraic connectivity starts at its minimum value $\approx \frac{1}{{2
N_1}}$ as predicted by (\ref{eq_1linkFirstOrderApprox}), which grows until it
reaches its maximum value $\mu_{N-1}(Q_1)$ when sufficient interlinks are
added. This means that a network with $100\%$ diagonal interlinks and the same
network with $90\%$ interlinks synchronize virtually at the same speed.
Comparing the final values of the algebraic connectivity, it is remarkable that
random networks synchronize faster than lattice networks. This is reasonably
due to the longer average distance in the latter.

The Fiedler cut starts at $\frac{1}{{2 L_1}}$ for a single added interlink.
Notice that it increases linearly with the percentage of interlinks, because
all added interlinks directly become part of the Fiedler cut. For all networks,
we observe a tipping point (which depends on the network type) upon which
adding a single link abruptly readjusts the partition: the Fiedler cut switches
from pure interlink cutting to a cutting of an invariable set of intralinks.
This abrupt change breaks the linearity.

The interdependence angle metric tells us that the Fiedler vector starts being
parallel to the first order approximation $x^{(0)}$ for $1$ added interlink.
Progressively, the Fiedler vector crawls the $N$-dimensional space up to the
transition point, where it abruptly jumps to the final (orthogonal) state
$\left(x_{N-1}(Q_1),x_{N-1}(Q_1)\right)$. Similarly to the interdependence
angle, the high values of entropy reflect the flatness of $x^{(0)}$, where all
components have (almost) the same absolute value. At this initial point,
entropy is maximum and almost equal to $log(2 N_{1})$, which tells us that the
initial partition consists purely of interlinks. When the partition turns to
the final state, the entropy is instantly shaped by the network topologies of
$A_1$ thus dropping to relatively much lower values. Notice that, for all
values of $N$, the highest final entropy is attained by the lattice graph due
to its regular structure, as seen in Fig.~\ref{fig_diagonal_results_average}.

\subsubsection{Network Model Differences}

RR and BA synchronize relatively faster than deterministic networks because
random interconnections shorten the average hopcount, thus bringing all
elements of the network closer (creating the \textit{small-world} effect
\cite{WatStr98}). For the particular case of BA, we observe the emergence of a
dominant partition which contains approximately $90\%$ of the total number of nodes.

There exists a significant difference between the $1,000$ node lattice and the
$10,000$ node lattice, which is expected due to the variable size response of
network models. We conjecture that this difference is caused by the average
geodesic distance:  the average node distance for a three dimensional lattice
lattice graphs grows with $\sqrt[3]{N}$, as opposed to random models, which
usually display logarithmic increases. Interestingly,
Fig.~\ref{fig_diagonal_results_average} shows that small lattices synchronize
faster than WS, but the situation is soon reversed for higher $N$.

To test whether the phase transition is merely an artifact of our synthetic
models, additional simulations were carried out using real topologies from the
KONECT dataset. Simulations verify that the transition from the natural
partition to the final orthogonal partition also occurs in real networks.
However, the transition takes place very early in the link addition process,
due to the poor synchronization capabilities of networks not designed for such
purpose. The interpretation of such result is that, to provide that real
network with a complete backup mirror without synchronization delays, a small
number of interlinks are required.

\subsection{General Interlinks Strategy}

\begin{figure*}[!ht]
\centering
\subfloat[Algebraic Connectivity]{
\includegraphics[width=0.45\textwidth]{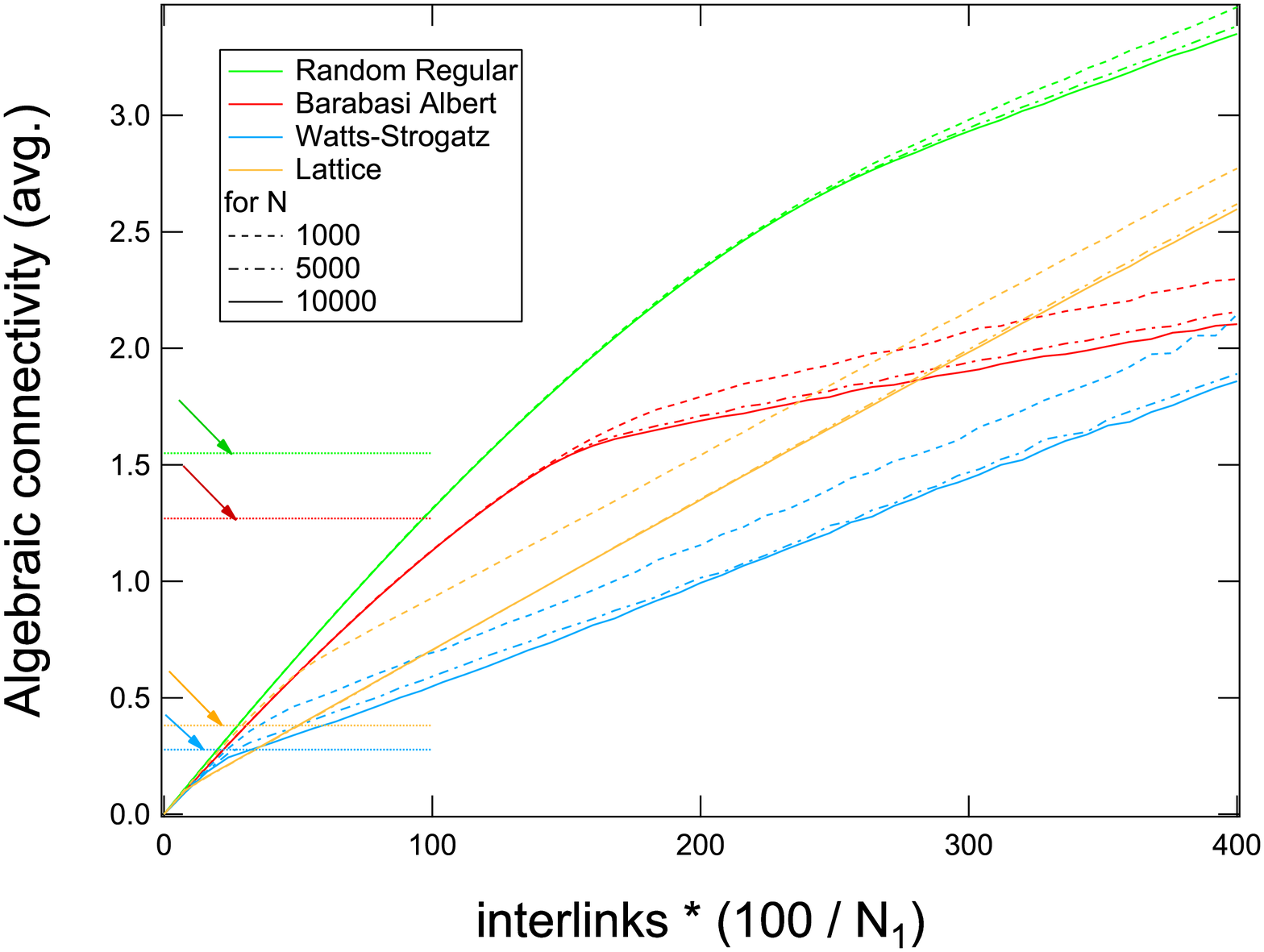}
\label{fig_general_results_average1}}
\subfloat[Fiedler cut]{
\includegraphics[width=0.45\textwidth]{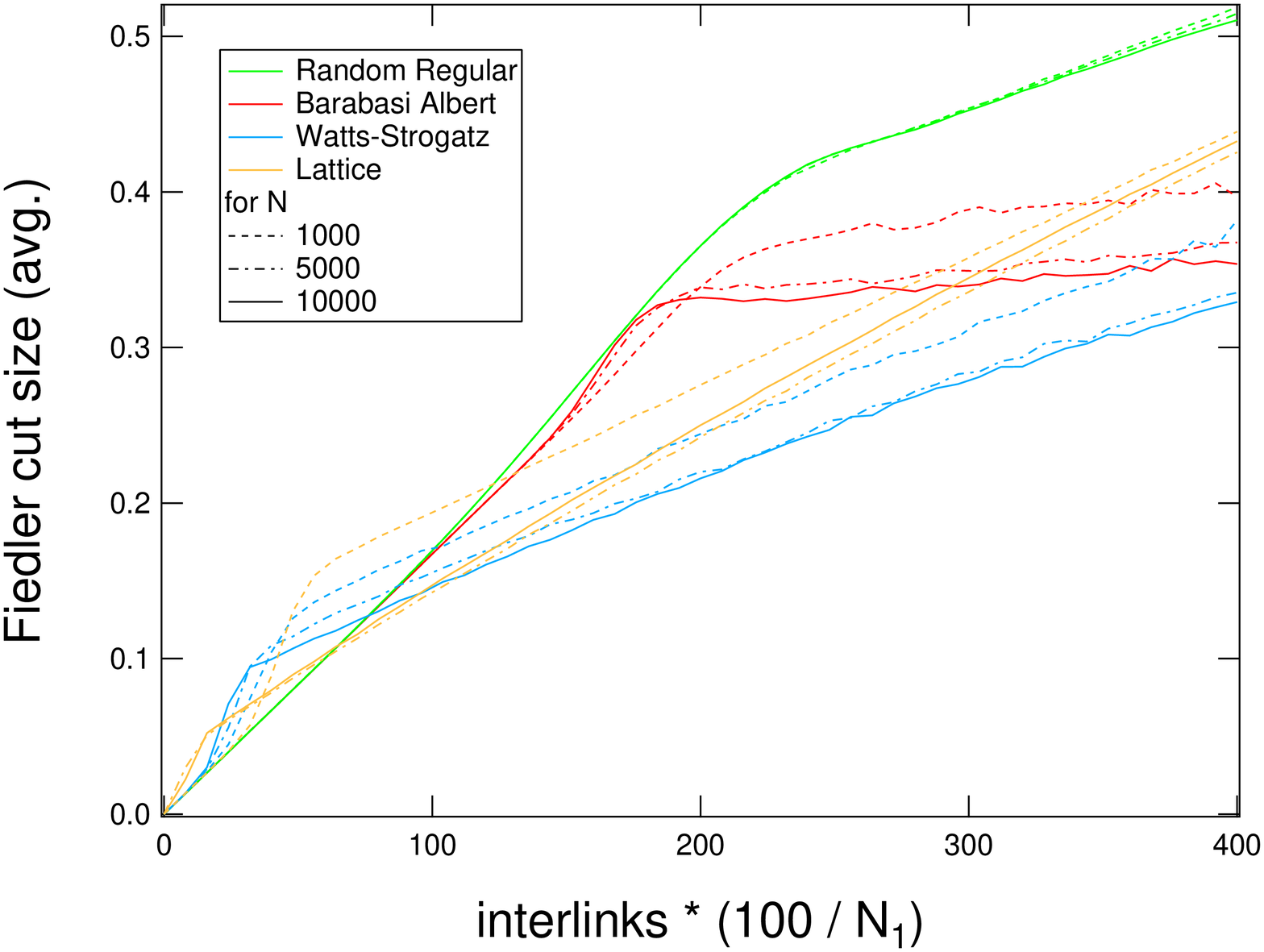}
\label{fig_general_results_average2}}
\\
\subfloat[Interdependence angle]{
\includegraphics[width=0.45\textwidth]{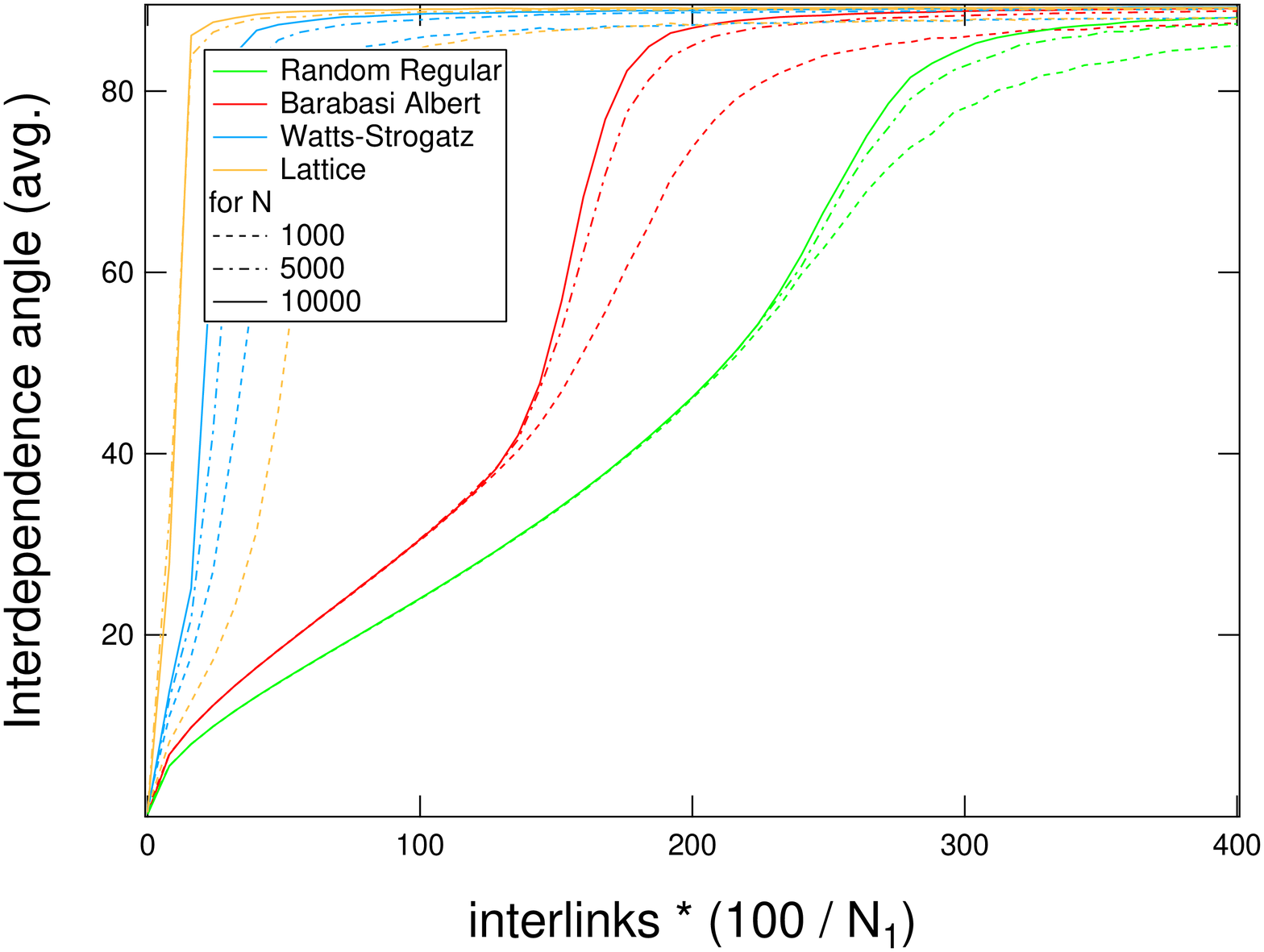}}
\subfloat[Entropy]{
\includegraphics[width=0.45\textwidth]{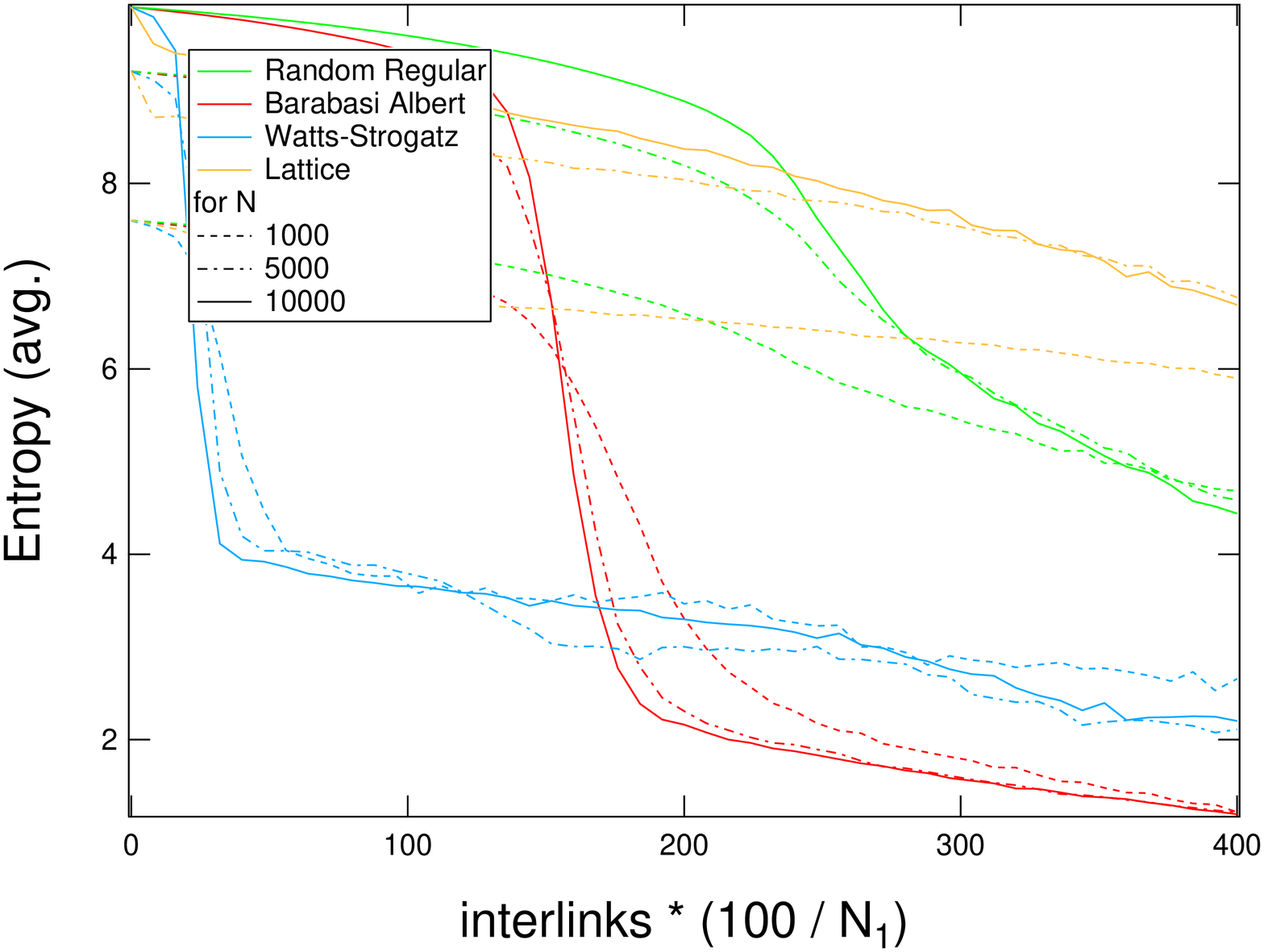}}
\caption{(Color online) Four metrics' averages are displayed to qualitatively
evaluate the effect of adding interlinks following the general strategy:
algebraic connectivity ($\mu_{N-1}$), Fiedler cut (${l\left(
{{{\mathcal{R}}},{{\mathcal{S}}}} \right)}/{{L_1} + {L_2}}$), interdependence
angle ($acos\left(x^T x^{(0)} / \left\| x \right\| \left\|x^{(0)}\right\|
\right)$), and entropy ($  - \sum\nolimits_{i = 1}^N {{x_i}^2\log {x_i}^2}$).
The four metrics indicate the existence of up to three regimes, but the regime
transitions are not as sharp as in the diagonal strategy scenario.
The flat lines signaled with arrows in the top left plot represent the average
algebraic connectivity of the $N_1=1,000$ respective single networks.}
\label{fig_general_results_average}
\end{figure*}
\begin{figure*}[!ht]
\centering

\subfloat[Fiedler cut]{
\includegraphics[width=0.45\textwidth]{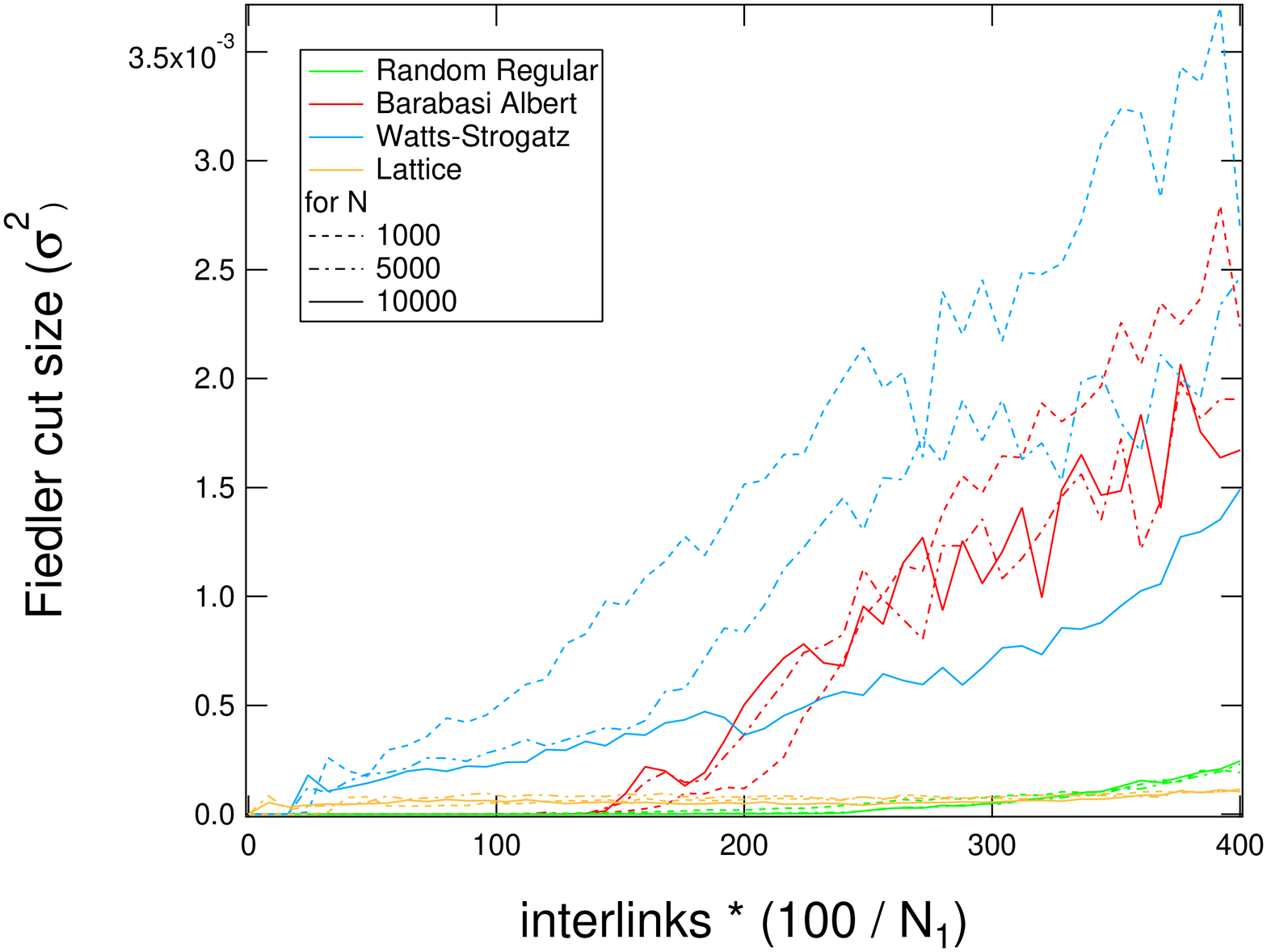}
\label{fig_general_results_stdeviation1}}
\subfloat[Interdependence angle]{
\includegraphics[width=0.45\textwidth]{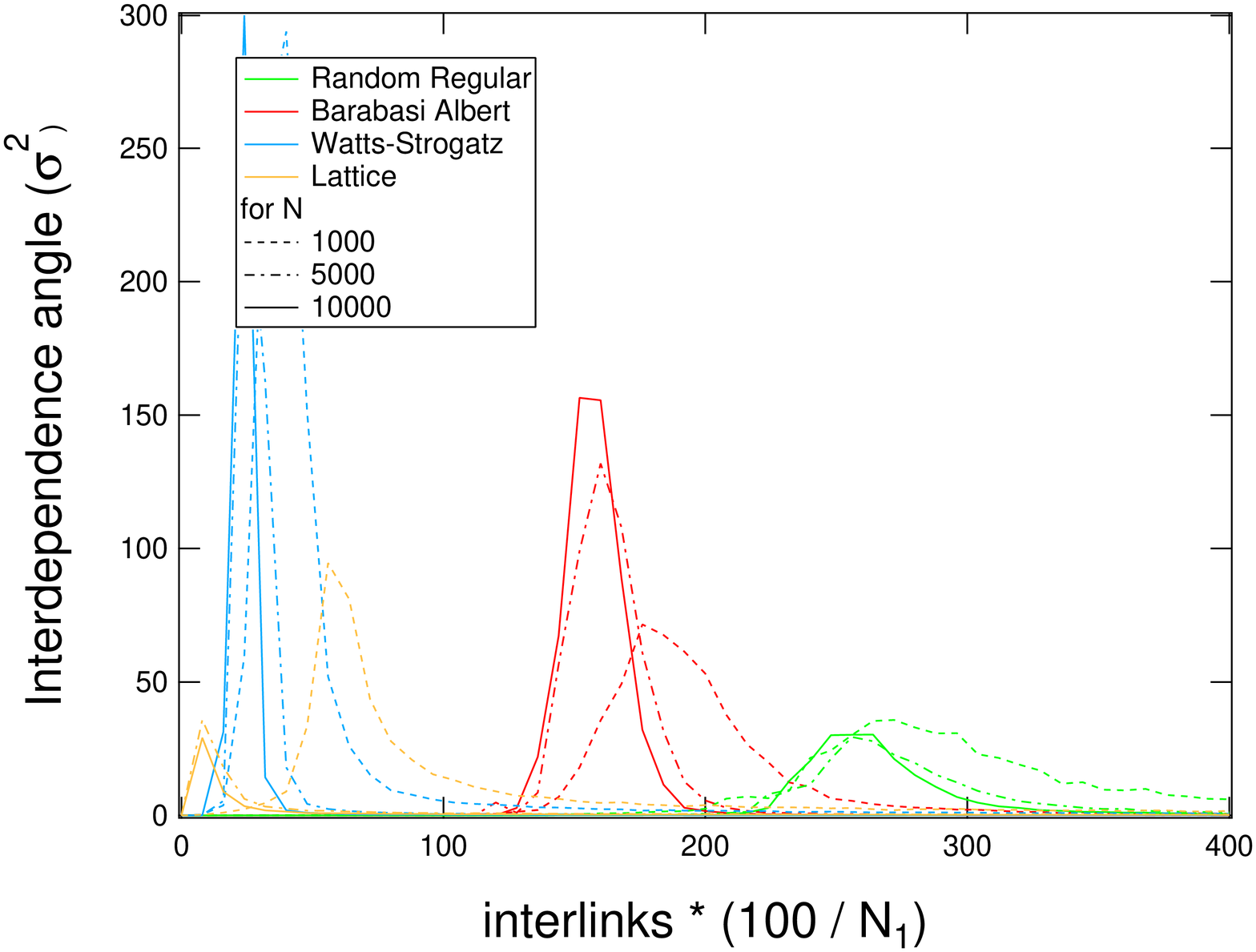}
\label{fig_general_results_stdeviation2}}
\caption{(Color online) The fluctuation $\sigma^2$ of the Fiedler cut and the
interdependence angle are displayed to evaluate the effect of adding interlinks
following the general strategy. The increasing Fiedler cut's fluctuations do
not hint the existence of a transition.
However the peaking fluctuation displayed by the interdependence angle
suggests the presence of a narrowing transition.}
\label{fig_general_results_stdeviation}
\end{figure*}

\subsubsection{Strategy Description}

As a variation of the localized diagonal interlinking strategy, our second
strategy randomly draws interlinks among any pair of nodes belonging to
different networks. Mean-field approximation provides us with exact results,
however perturbation analysis loses its power when too many links are added,
i.e. the perturbation can no longer be regarded as small. For this reason, we
cannot predict an exact asymptotic state as for the diagonal strategy. We have
limited our simulations to the inclusion of up to $4$ interlinks per node.

\subsubsection{Effect on partition quality metrics}

We can observe that the algebraic connectivity of all models experiences two
regimes, upon the progressive addition of interlinks as illustrated in
Fig.~\ref{fig_general_results_average}. Initially, for a small number of added
links, the initial state dips to a minimum as is the case for the diagonal
strategy and represents a good starting point for the perturbation theory. As
we increase the number of interlinks, the average algebraic connectivity and
Fiedler cut curves show a linear increase. At the critical number of links
$l_{J} = \mu_{N-1} \cdot N$, the average slope switches regime by damping to
half its value, as seen in Fig.\ref{fig_general_results_average1} and
Fig.~\ref{fig_general_results_average2}, which is in perfect agreement with our
theoretical prediction (\ref{eq_generalCase}). However not only the average,
but also the fluctuations steadily increase, as illustrated by
Fig.~\ref{fig_general_results_stdeviation1}. High fluctuations are expected,
due to the large set of available graph configurations.

As we can see from the interdependence angle in
Fig.~{\ref{fig_general_results_average}}, in the first regime the natural
partition is partially preserved up to $l_{J}$. The interdependence angle
experiences a sharp increase at the turning point, which further narrows as $N$
increases as seen in Fig.~\ref{fig_general_results_stdeviation2}. This is due
to the fact that the Fiedler cut in all our isolated model networks scales less
than linearly with the network size, which is consistent with the picture of a
phase transition between a Fiedler cut dominated by interlinks and an other
dominated by intralinks. As opposed to the diagonal strategy, the final
eigenvector is not strictly identical to the Fiedler eigenvector of the
isolated networks $x_{N-1}$, but it also involves interlink cuts. This is due
to the fact that in the general case $x_{N-1}$ does not belong to the kernel of
$Q_B$ as opposed to the diagonal case.

The exact location of the phase transition can also be predicted employing
perturbation theory, by imposing the perturbed value $x_{N-1}(Q)$ of the
configuration to be equal to that achieved starting from the $x_{N-1}(Q_1)$
initial state. However, the resulting formulas are not particularly simple and
their numerical calculation requires a time comparable with the Fiedler
eigenvalue evaluation of the sparse metrics. For this reasons such estimates
are not reported here.

\subsubsection{Network Model Differences}

Let us focus on the case of adding a small number of interlinks in the range
$\left[1,N\right]$. The diagonal strategy will synchronize faster than the
general strategy in the case of RR and BA, as illustrated in
Fig.~\ref{fig_general_results_average1}. On the other hand, the general
strategy synchronizes faster in WS and LA models. Thus if we were to add
precisely $N_1$ general interlinks between two identical networks, regular
structures would (relatively) benefit the most.

For BA, the fraction of intralinks belonging to the Fiedler partition decreases
with increasing number of interlinks, whereas the
$\frac{\mathcal{R}}{\mathcal{S}}$ ratio increases. This hints that nodes group
into high degree clusters (with a high link/node ratio) and a low degree
clusters (with a low link/node ratio). In addition, BA's entropy experiences
the highest drop, which indicates that the Fiedler vector is highly localized
around a small set of nodes.

The difference between random and grid networks still exists for the general
strategy, but it is not as predominant as in the diagonal case. This effect is
expected due to the randomization resulting from the random addition of links
to regular structures, which is the conceptual basis of the WS model. In
general, we observe that the optimal link addition strategy depends on the
network topology.

\section{Conclusions} \label{sec_concl}

This paper aims to provide general results concerning the synchronization of
interdependent identical networks. We provided evidence that upon increasing
the number of interlinks between two originally isolated networks, their
synchronizability experiences a phase transition. That is, there exists a
critical number of \textit{diagonal interlinks} beyond which any further
inclusion does not enhance synchronization capabilities at all. Similarly,
there exists a critical number of \textit{general interlinks} beyond which
algebraic connectivity increments at half the original rate.

The exact location of the transition depends exclusively on the algebraic
connectivity of the graph models, and it is always observed regardless of the
interconnected graphs. For the two proposed interconnection strategies, the
critical number of interlinks that triggers the transitions can be predicted
correctly by mean-field approximations : $\frac{\mu_{N-1}(Q_1) \cdot N_{1}}{2}$
links for the \textit{diagonal interlinks} strategy, and $\mu_{N-1}(Q_1) \cdot
N_{1}$ links for the \textit{general interlinks} strategy. By resorting to
perturbation theory we have provided upper bounds for the total algebraic
connectivity of the interdependent system and means to estimate it.

This paper beacons a significant starting point to the understanding of the
mutual networks synchronization phenomena, as we have just started studying
this extremely interesting field. Nonetheless, different linking strategies
should be researched and general theory developed. Regarding the mutual
synchronization of heterogeneous networks (i.e. $A_1 \neq A_2$), preliminary
results confirm the existence of phase transitions with similar features to the
general random linkage of identical networks. However, we could not observe any
dominant strategy as in the case with the diagonal interlinking.

\subsection*{Acknowledgements}

This research has been partly supported by the European project MOTIA (Grant
JLS-2009-CIPS-AG-C1-016); the EU Research Framework Programme 7 via the CONGAS
project (Grant FP7-ICT 317672); and the EU Network of Excellence EINS (Grant
FP7-ICT 288021).

\end{document}